\documentclass[aps,prapplied,twocolumn,superscriptaddress,longbibliography]{revtex4-1}

\usepackage{color}
\usepackage{setspace}
\usepackage{graphicx}
\usepackage{mathtools}
\usepackage{verbatim}
\usepackage{amsmath}
\usepackage{caption}
\usepackage{appendix}
\usepackage{esint}
\usepackage{commath}
\usepackage{tensor}
\usepackage{stackengine}
\usepackage{xcolor,soul}
\usepackage{mathrsfs}
\usepackage{dcolumn}
\usepackage{bm}
\usepackage[english]{babel}

\begin{document}

\title{The Electrodynamics of Free and Bound Charge Electricity Generators using Impressed Sources and the Modification to Maxwell's Equations}

\author{Michael E. Tobar}
\email[]{michael.tobar@uwa.edu.au}
\author{Ben T. McAllister}
\author{Maxim Goryachev}
\affiliation{ARC Centre of Excellence for Engineered Quantum Systems and ARC Centre of Excellence for Dark Matter Particle Physics, Department of Physics, University of Western Australia, 35 Stirling Highway, Crawley, WA 6009, Australia.}

\date{\today}

\begin{abstract}
{Electric generators convert external energy, such as mechanical, thermal, nuclear, chemical and so forth, into electricity and are the foundation of power station and energy harvesting operation. Inevitably, the external source supplies a force per unit charge  (commonly referred to as an impressed electric field) to free or bound charge, which produces AC electricity. In general, the external impressed force acts outside Maxwell's equations and supplies a non-conservative electric action generating an oscillating electrodynamic degree of freedom. In this work we analyze the electrodynamics of ideal free and bound charge electricity generators by introducing a time dependent permanent polarization, which exists without any applied electric field, necessarily modifying the constitutive relations and essential to oscillate free or bound charge in a lossless way. For both cases, we show that Maxwell's equations, and in particular Faraday's law are modified, along with  the required boundary conditions through the addition of an effective impressed magnetic current boundary source and the impressed electric field, related via the left hand rule. For the free charge case, we highlight the example of an electromagnetic generator based on Lorentz force, where the impressed force per unit charge that polarizes the conductor, comes from mechanical motion of free electrons due to the impressed velocity of the conductor relative to a stationary DC magnetic field. In contrast, the bound charge generator is simply an idealized permanently polarized bar electret, where the general case of a time dependent polarized electret is the underlying principle behind piezoelectric nano-generators. In the open circuit state, both bound and free charge electricity generators are equivalent to idealized Hertzian dipoles, with the open circuit voltage equal to the induced electromotive force (emf). Analyzing the short circuit responses, we show that the bound charge electricity generator has a capacitive source impedance. In contrast, we show for the ideal free charge AC electricity generator, the back emf from the inductance of the loop that defines the short circuit, directly cancels the source emf, so the voltage across the inductor is solely determined by the magnetic current boundary source. Thus, we determine the magnetic current boundary source best describes the output voltage of an AC generator, rather than the electric field.}
\end{abstract}


\maketitle

\section{Introduction}

The most common form of electricity generation converts motive (or mechanical) energy into photonic or electromagnetic energy described by Faraday's law and is known as an alternating current (AC) induction generator.  In general, the creation of photons or electricity via Faraday's law must come from another external energy source to drive the turbines into motion. The conservation of energy is a fundamental law of physics, but only applies to an isolated system. This law means that energy cannot be created or destroyed, but only transferred from one form to another. For example, a nuclear reaction where mass is converted into other forms of energy through $E=mc^2$. In this case, if the other form of energy is electricity, then we can call this device a nuclear generator or battery. For example, a nuclear battery uses the energy from the decay of a radioactive isotope to generate electricity and can produces large DC electric fields and voltages of up to 10-100 kV \cite{Coleman53}. The modern form of the nuclear battery is a micro-electromechanical system or MEMS device \cite{NucBatt,nucbat2002}, which can also be configured as an AC generator capable of generating radio frequencies of $60-260 MHz$ \cite{Li2001}. A more recent type of generator is the nanogenerator, which uses special materials that convert an external strain or motion to a time-dependent electric polarization of a material independent of an applied electric field\cite{WANG20179,WANG201774}. Such a device can convert mechanical energy to electricity via a piezoelectric or triboelectric effect \cite{Sessler2016,Wang242,Yang:2009wa,FAN2012328,Wang2013}, or temperature variations to electricity via a pyroelectric effect \cite{XUE2017147,Yang20122833,Ko20166504,Yang20125357,Zi20152340,Lee2014765,Park201511830}.

When considering the generation of electricity only from the view point of the created electromagnetic degree of freedom, the creation of electromagnetic energy is a non-conservative process. Thus, when we consider the electrodynamics in isolation to the whole system, the standard Maxwell's equations must be made more general to take into account the non-conservative processes. This requires the ability to add the impressed forces into Maxwell's equations, and we show in this work, that this can be achieved by generalising the constitutive relations. The modification of the constitutive relations inevitably leads to a modification of Maxwell's equations, in a similar way to what happens when we consider the difference between Maxwell's equations in Matter and in Vacuum. For example, in a dielectric medium, the applied electric field polarizes the material, which causes an opposing internal electric force, which reduces the electric field in the medium, which modifies Gauss' Law. 

The standard Maxwell's equation in differential form and in dielectric and magnetic media are in general given by (SI units),
\begin{align}
&\vec{\nabla}\cdot\vec{D}=\rho_f,\label{V1}
\end{align}
\begin{align}
&\vec{\nabla} \times \vec{H}-\frac{\partial \vec{D}}{\partial t}=\vec{J}_f,\label{V2}
\end{align}
\begin{align}
&\vec{\nabla} \cdot \vec{B}=0,\label{V3}
\end{align}
\begin{align}
&\vec{\nabla} \times \vec{E}+\frac{\partial \vec{B}}{\partial t}=0, \label{V4}
\end{align}
where
\begin{align}
&\vec{D}=\epsilon_0\vec{E}+\vec{P}
&\vec{H}=\mu_0^{-1}\vec{B}-\vec{M}, \label{Aux}
\end{align}
and
\begin{align}
&\rho_b=-\nabla\cdot\vec{P}
&\vec{J}_b=\nabla\times\vec{M}, \label{Bound}
\end{align}
Here $\vec{E}$ is the electric field intensity, $\vec{H}$ is the magnetic field intensity, $\vec{D}$ is the electric flux density, $\vec{B}$ is the magnetic flux density, $\vec{J}_f$ is the free electric current density, $\vec{J}_b$ is the bound electric current density, $\rho_f$ is the free electric charge density, $\rho_b$  is the bound electric charge density and $\epsilon_0$ and $\mu_0$ are the permittivity and permeability of free space. 

When an electric field is applied to a dielectric material its molecules respond by forming microscopic electric dipoles, which causes a distribution of bound charge dependent on the applied field $\vec{E}$, which acts to reduce the electric field in the dielectric. For a dielectric material the macroscopic polarization vector, $\vec{P}$ is related to the average microscopic bound charge density by eqn. (\ref{Bound}). In a similar way when a $\vec{B}$-field is supplied to a magnetic material, magnetic moments of the atoms align to cause a macroscopic magnetization $\vec{M}$ related to the bound current given in eqn. (\ref{Bound}). Thus in matter, Maxwell's equations can be fully specified in terms of the $\vec{E}$ and $\vec{B}$ fields along with the auxiliary fields $\vec{D}$ and $\vec{H}$ given by eqns. (\ref{V4}), combined with the constitutive relations given by eqns. (\ref{Aux}). However, in order to fully apply these equations the general relationships between the fields in eqns. (\ref{Aux}) must be further specified, which depend on the material properties. In general this can be quite complex, as materials may be anisotropic, magneto-electric, piezoelectric, and so forth. The simplest form of the constitutive relations represents isotropic and linear materials such that $\vec{D}=\epsilon_{0}\epsilon_{r}\vec{E}$ and $\vec{H}=\mu_0^{-1}\mu_r^{-1}\vec{B}$, where $\vec{P}=\epsilon_{0}\chi_{e0}\vec{E}$ so $\epsilon_{r}=1+\chi_{e0}$ and $\vec{M}=\chi_{m0}\vec{H}$  so $\mu_r=1+\chi_{m0}$. The free and bound currents and charges in Maxwell's equations given above are not source terms for the fields and do not add energy into the system. They either propagate without loss due to the interaction with the electromagnetic fields, or can describe a dissipative (or resistive) system where electromagnetic energy is lost, usually by conversion to heat (in this case the electric and magnetic field phasors can become complex).

The non-electric energy sources (for example, nuclear energy as discussed previously) capable of transmitting energy and hence a force to electric charges are commonly referred to as ``impressed" sources of the field. Theoretically, they can be represented either as an ideal current or voltage generator of electronic network theory \cite{Popovic81}. Thus in electrodynamics, an impressed electric current needs to be added as a current source and hence will modify Ampere's law. In contrast, we show an impressed electric voltage needs to be added as a magnetic current source, and hence will modify Faraday's law. The later does not mean that magnetic monopole particles exists, but is a consistent way to model boundary value problems when considering the electrodynamics of a non-conservative electricity generator \cite{RHbook2012,Balanis2012,ECJ}. This technique is more generally known as the Compensation Theorem \cite{Popovic81,Monteath51,ERAntTh18}. We also note the impressed sources are not influenced by Maxwell's equations because they represent creation of electromagnetic energy from an external source. Two-potential theory is summarized in the appendix, which is a common way to include the non-conservative impressed terms in antenna theory.

In this paper we analyse the electrodynamics of free and bound charge electricity generators using impressed sources. Importantly, we have shown that the impressed sources modify the constitutive relations, which essentially are the same as the force balance equations between the external energy source and the electrodynamic generated degree of freedom. We show that this modifies Maxwell's equations with the addition of a free or bound charge macroscopic polarization, created without an applied electric field. Because the vector curl of this polarization is non-zero, we have shown that Faraday's law must be modified. We then apply this technique to explain the physics of some free charge and bound charge electricity generators. In particular we find that the definition of the voltage, $V_e$, of the electricity generator is best described by, $V_e=-I^i_m$, where $I^i_m$ is the total effective impressed magnetic current at or near the boundary. This approach is similar to the approach recently undertaken to explain axion modifications to electrodynamics, where the axion mixes with a photon to acts as the external impressed force, which converts axions into electricity \cite{TobarModAx19,TOBAR2020100624,Cao2017}. 

\section{Electrodynamics of the Generation of Electricity from Free Charge}

\begin{figure}[t]
\includegraphics[width=1.0\columnwidth]{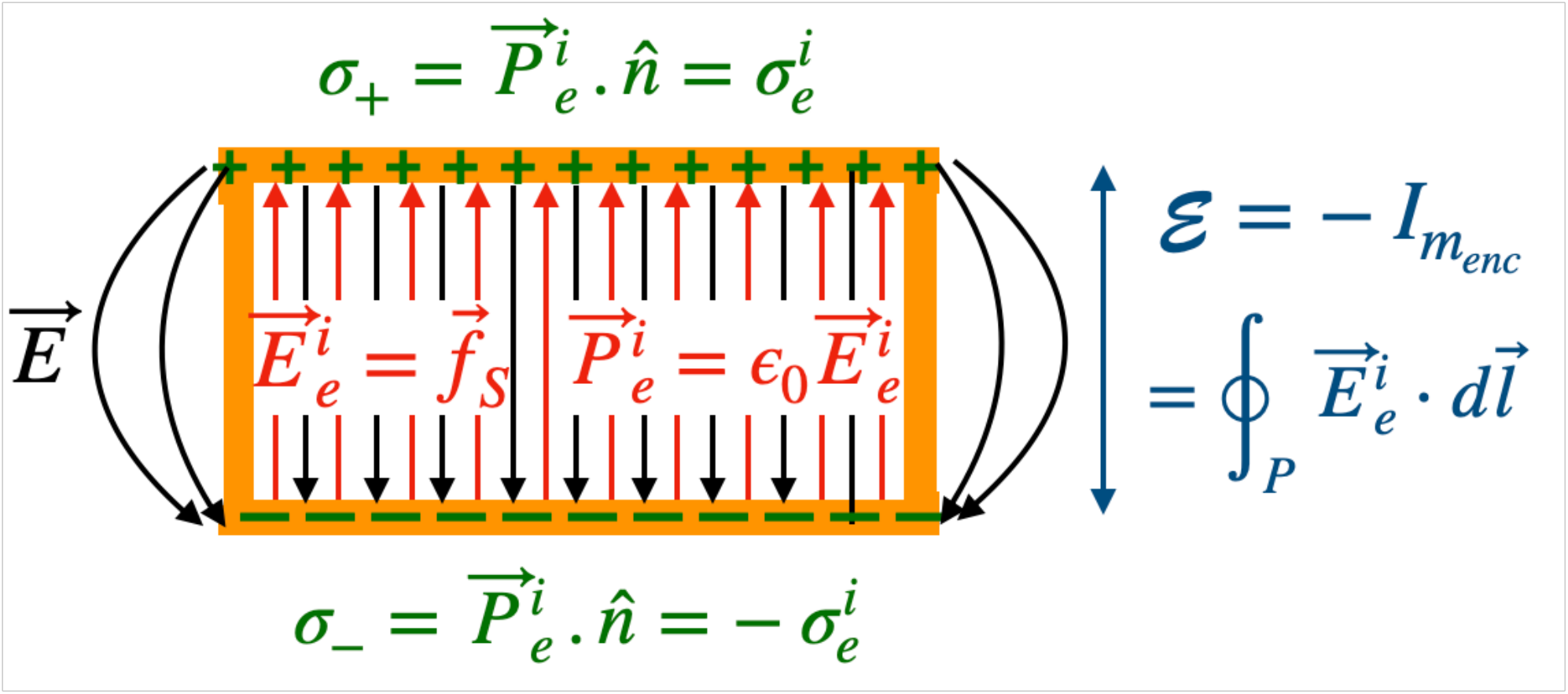}
\caption{Illustration of the electric field generated in a free-charge ideal DC voltage source from an external energy source. The associated delivered external force per unit charge, $\vec{f}_{S}$, supplies the energy to seperate (and hence polarize) impressed free surface charges, $\sigma^i_{e}$ at the axial boundaries. This is equivalent to an impressed electric field, $\vec{E}^i_e=\vec{f}_{S}$, or polarization, $\vec{P}^i_e$, as given by equations (\ref{freeE}) and (\ref{freeE2}), and related to a magnetic current at the radial boundary, $I_{m_{enc}}$, by the left hand rule. The non-conservative nature means an electromotive force as calculate in eqn. (\ref{Im}), $\mathcal{E}=-I_{m_{enc}}=\oint_P\vec{E}^i_e\cdot d\vec{l}$, is generated, resulting in a voltage output that can drive an electric circuit. The separated free charges then generate a conservative  dipole electric field, $\vec{E}$.}
\label{eleV}
\end{figure}

To analyse a free charge voltage generator at DC, or in the quasi-static limit, we can start with the equations given in advanced electrodynamics text books such as Griffiths \cite{GriffBook}, where he shows that the total force per unit charge, $\vec{f}$ involved in a free charge DC voltage source is given by,
\begin{equation}
\vec{f}=\vec{E}+\vec{f}_S.
\label{freeE}
\end{equation}
Here $\vec{f}_S$ is the force per unit charge, which supplies the energy to seperate the charges and supply an electromotive force (emf) from an external energy source. Following this a resulting electric field, $\vec{E}$, is produced by the separated charges. Harrington \cite{RHbook2012} presents essentially the same equation as Griffiths \cite{GriffBook} for a general AC generator, but using different terminology. Harrington considers the electric field more generally such that a total field, $\vec{E}_T \equiv \vec{f}$, which is consistent with the $\vec{E}_T$ in the appendix. Harrington also defines $\vec{E}^i_e \equiv \vec{f}_S$ as the source impressed electric field. Here to be consistent with Griffith \cite{GriffBook} and the left hand rule for the relation between magnetic current and electric field, we have defined $\vec{E}^i_e$ in the opposite direction to Harrington, so that
\begin{equation}
\vec{E}_T=\vec{E}+\vec{E}^i_e,
\label{freeE2}
\end{equation}
The effective impressed field, $\vec{E}_e^i$, is confined to the voltage source and does not exist outside it. 

In a DC battery where external energy is converted to electromagnetic energy \cite{Roberts1983,Baierlein2001,Saslow1999}, the impressed electric field allows the electrons to move in the opposite direction to the electric field created by the electrons themselves, even though the internal resistance inside an ideal free charge voltage source is near zero, with (ignoring fringing) $\vec{E}\approx-\vec{E}^i_e$ and thus $\vec{E}_T\approx 0$ (these values depend highly on aspect ratio and hence the fringing field). An example of such a voltage source is illustrated in Fig.\ref{eleV}. Given that the impressed force per unit charge $\vec{E}^i_e$ is non-conservative, there is an effective impressed magnetic current boundary source linked by the closed path, as shown in Harrington and Balanis \cite{RHbook2012,Balanis2012}, which means for the DC case,
\begin{equation}
\vec{\nabla} \times\vec{E}_e^i=-\vec{J}_m^i,~~\vec{\nabla} \times\vec{E}=0,~~\vec{\nabla} \times\vec{E}_T=-\vec{J}_m^i.
\label{fara}
\end{equation}
Then by Stokes' theorem we can calculate the emf of the voltage generator, $\mathcal{E}$, by,
\begin{align}
&\mathcal{E}=-I_{m_{enc}}^i=\oint_P\vec{E}_e^i\cdot d\vec{l}=\oint_P\vec{E}_T\cdot d\vec{l},
\label{emff}
\end{align}
where the path encloses the effective impressed magnetic current at the radial boundary, given by, 
\begin{align}
I_{m_{enc}}^i=\int_S\vec{J}_m^i\cdot d\vec{a}. 
\label{Im}
\end{align}
Since $\vec{E}_e^i$ will apply a force per unit charge to seperate free charge in the system, an impressed free charge distribution, $\rho_e^i$ will be created so that,
\begin{equation}
\epsilon_0\nabla\cdot \vec{E}_e^i=\nabla\cdot \vec{P}_e^i=-\rho_e^i,
\label{modGauss}
\end{equation}
effectively polarizing the free charge (creating a dipole), and allowing the definition of a permanent free charge polarization of, $\vec{P}^i_e=\epsilon_0\vec{E}^i_e$. For the situation in Fig.\ref{eleV}, the impressed free charges occur as surface charge at the ends of the DC voltage source, given by,
\begin{equation}
\sigma_\pm=\vec{P}^i_e.\hat{n}=\pm\sigma_e^i.
\end{equation}
According to this definition, the ideal DC voltage source will only have impressed free charge, and thus $\epsilon_0\nabla\cdot \vec{E}=\rho_e^i$, as a result. The net result is the system has both an electric vector and scalar potential and outside the voltage source the electric field resembles a capacitor like dipole even though it is modelled as a perfect conductor. The next step is to expand this technique to an AC electricity generator.

\subsection{Time-Dependent Free Charge Electricity Generation}

The DC system described previously has no magnetic field component in the system and adding time dependence will potentially induce a magnetic field. First, we note that the impressed magnetic current boundary source is divergenceless. This means from two potential theory in the appendix, $\rho^i_m=0$ and $\vec{\nabla}\cdot\vec{B}=0$. Following this we can implement the quasi-static approximation and calculate the $\vec{B}$-field from the modified Ampere's law. To calculate this, we combine with the continuity equation in eqn. (\ref{J1}) for the impressed free current, $\vec{J}^i_{e}$, with eqn. (\ref{modGauss}) to obtain, 
\begin{equation}
\vec{J}^i_{e}=\frac{\partial\vec{P}_e^i}{\partial t}=\epsilon_0\frac{\partial\vec{E}_e^i}{\partial t}.
\end{equation}
Assuming a lossless system then any other free current besides the impressed current, $\vec{J}^i_{e}$, must be divergence free, so $\nabla\cdot\vec{J}_{f}=0$ and $\rho_f=0$ in eqn. (\ref{E1}) and (\ref{E2}) in the appendix, then in the time varying case, the modified Ampere's law is given by, $\vec{\nabla} \times \vec{B}-\epsilon_0\mu_0\frac{\partial \vec{E}_T}{\partial t}=\vec{J}_{f}$, with a modified Gauss' law of $\vec{\nabla} \cdot \vec{E}_T=0$. The next step of the quasi-static approximation is to calculate the created $\vec{E}$-field from the calculated $\vec{B}$-field from Faraday's law, $\nabla\times\vec{E}+ \frac{\partial\vec{B}}{\partial t}=0$, then combining with eqn. (\ref{fara}), the modified Faraday's law becomes, $\vec{\nabla} \times \vec{E}_T+\frac{\partial \vec{B}}{\partial t}=-\vec{J}_m^i$, here the $\frac{\partial \vec{B}}{\partial t}$ term can be considered as a displacement magnetic current. The only part of Maxwell's equations, which remains unmodified in this case is the magnetic Gauss' law.  

Thus, Maxwell's equations in differential form may be written in terms of $\vec{E}_e^i$ and $\vec{E}$, by,
\begin{align}
&\vec{\nabla}\cdot\vec{E}=\frac{\rho_e^i}{\epsilon_0}~~\text{and}~~\vec{\nabla}\cdot\vec{E}_e^i=-\frac{\rho_e^i}{\epsilon_0},\label{Vs1}\\
&\vec{\nabla} \times \vec{B}-\epsilon_0\mu_0\frac{\partial \vec{E}}{\partial t}=\mu_0\vec{J}_e^i=\epsilon_0\mu_0\frac{\partial\vec{E}_e^i}{\partial t},\label{Vs2}\\
&\vec{\nabla} \cdot \vec{B}=0,\label{Vs3}\\
&\vec{\nabla} \times \vec{E}+\frac{\partial \vec{B}}{\partial t}=0~~\text{and}~~\vec{\nabla} \times \vec{E}_e^i=-\vec{J}_m^i. \label{Vs4}
\end{align}
or $\vec{E}_T$ by
\begin{align}
&\vec{\nabla}\cdot\vec{E}_T=0,\label{Vss1}\\
&\vec{\nabla} \times \vec{B}-\epsilon_0\mu_0\frac{\partial \vec{E}_T}{\partial t}=0,\label{Vss2}\\
&\vec{\nabla} \cdot \vec{B}=0,\label{Vss3}\\
&\vec{\nabla} \times \vec{E}_T+\frac{\partial \vec{B}}{\partial t}=-\vec{J}_m^i. \label{Vss4}
\end{align}
Note, we set $\vec{J}_f=0$, for the open circuit no-load case. Following this, the integral forms of Maxwell's equations may be written as,
\begin{equation}
\oiint_S\vec{E}\cdot d\vec{a} =\frac{Q^i_{e_{enc}}}{\epsilon_0}~~\text{and}~~\oiint_S\vec{E}_e^i\cdot d\vec{a} =-\frac{Q^i_{e_{enc}}}{\epsilon_0} \label{eq:V1},
\end{equation}
\begin{equation}
\oint_P \vec{B}\cdot d\vec{l}-\mu_0\epsilon_0\frac{d}{dt}\int_S(\vec{E}+\vec{E}_e^i)\cdot d\vec{a}=0
\label{eq:V2}
\end{equation}
\begin{equation}
\oiint_S \vec{B}\cdot d\vec{a}=0\label{eq:V3},
\end{equation}
\begin{equation}
\oint_P\vec{E}\cdot d\vec{l}+\frac{d}{dt}\int_S\vec{B} \cdot d\vec{a}=0~~\text{and}~~\oint_P\vec{E}_e^i\cdot d\vec{l}=-I_{m_{enc}}^i\label{eq:V4}
\end{equation}
We can also write eqn. (\ref{eq:V4}) as,
\begin{equation}
\oint_P\vec{E}_T\cdot d\vec{l}+\frac{d}{dt}\int_S\vec{B} \cdot d\vec{a}=-I_{m_{enc}}^i\label{eq:V5}
\end{equation}
and (\ref{eq:V1}) as
\begin{equation}
\oiint_S\vec{E}_T\cdot d\vec{a} =0 \label{eq:V1b},
\end{equation}

In general from a circuit theory perspective, the structures under consideration are considered as electrically small (quasi-static limit), no time delay exists between sources and the rest of the circuit and the only loss occurs through dissipation. From an antenna theory perspective these assumptions in general can be relaxed as time delays may be important. In this work we just consider the quasi-static limit relevant for antenna and circuit theory in the limit that the structures are small compared to the wavelength.

\subsection{Ideal Cylindrical AC Free Charge Electricity Generator}

\begin{figure}[t]
\includegraphics[width=1.0\columnwidth]{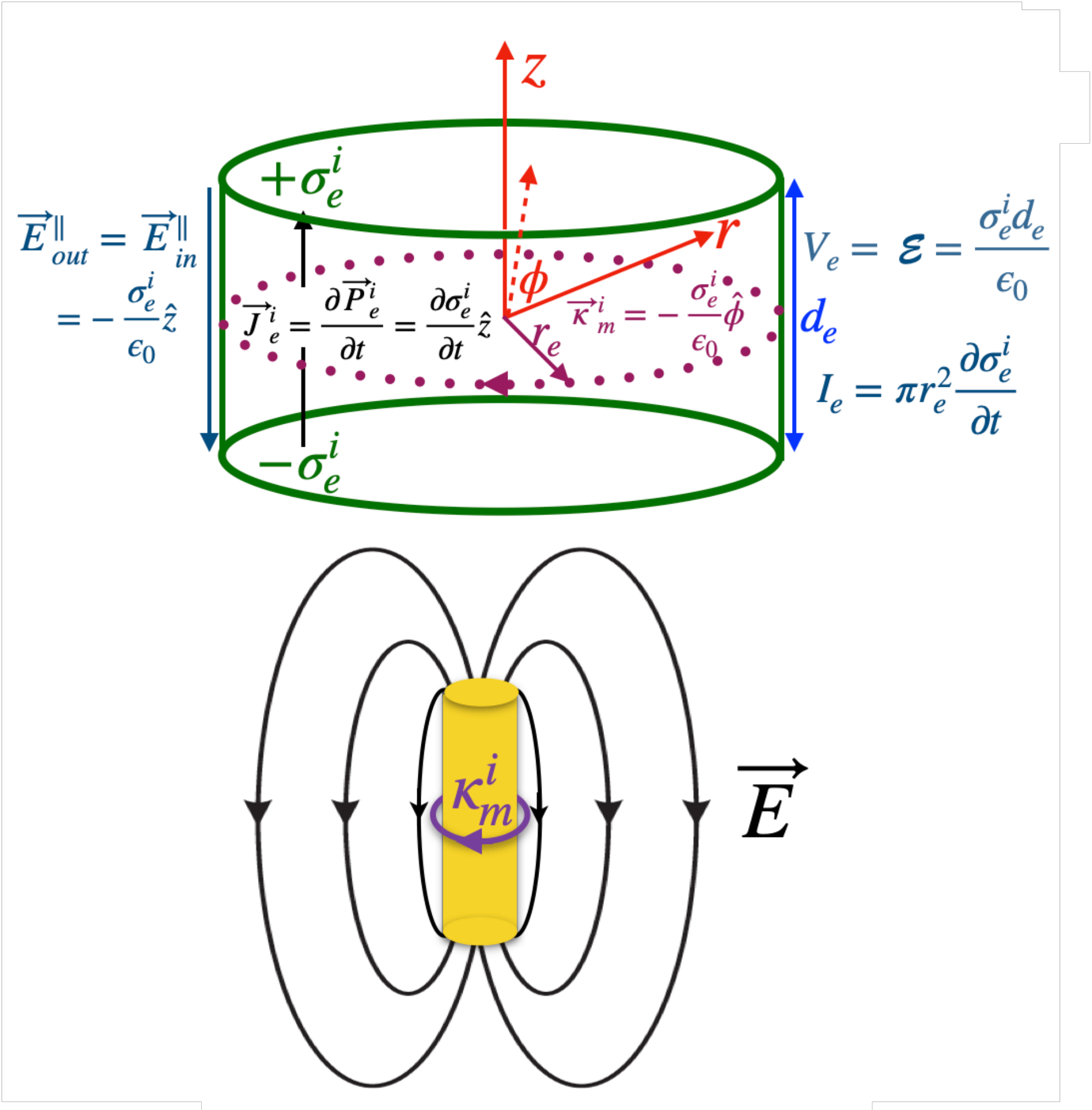}
\caption{Above, schematic of the cylindrical ideal free charge AC electric generator of axial length, $d_e$ and radius $r_e$, with associated field and source terms. For the ideal AC generator of terminal voltage, $V_e$, the impressed electric current density, $\vec{J}_e^i$, flows up and down the cylindrical axis as the terminal charges, $\pm\sigma_{e}^i$, oscillate in polarity. Below, the active near field of a Hertzian dipole antenna resembles that of a fringing field in a capacitor\cite{Hum20}, and is equivalent to the above free charge AC electric generator.}
\label{elVcyl}
\end{figure}

In the following we will analyse the electrodynamics of an ideal zero resistance cylindrical AC voltage generator, as shown in Fig.\ref{elVcyl}. We recognise that the net charge in the system is zero, and the charge will be present as a surface charge density, $\pm\sigma^i_{e}$, at the upper and lower boundaries of the cylinder. Furthermore, there will be a non-familiar boundary condition to be determined at the radial boundary, since  $\vec{E}_e^i$ must  be contained within the cylinder. 

For an ideal source in the quasi-static limit, we ignore any source impedance, including inductance or capacitance, so the AC emf is similar to the DC case (eqn. \ref{emff}), given by,
\begin{align}
&\mathcal{E}=-I_{m_{enc}}^i=\oint_P\vec{E}_e^i\cdot d\vec{l},
\label{emfff}
\end{align}
and the surface charge on the end faces caused by the impressed field, $\vec{E}_e^i$, can be calculated to be,
\begin{align}
&\sigma_e^i=\vec{P}_e^i\cdot\hat{n}=\epsilon_0\vec{E}_e^i\cdot\hat{n}, \ \text{where} \  \vec{P}_e^i =\sigma_e^i\hat{z}.\label{fEL9}
\end{align}
Here $\hat{n}$ is the normal to the surface, which is equal to $\hat{z}$ on the top surface and -$\hat{z}$ on the bottom surface. Then, from eqn.(\ref{emff}) the emf generated in the quasi-static limit is calculated to be,
\begin{align}
&\mathcal{E}=E_e^id_e=\frac{\sigma_e^id_e}{\epsilon_0}=V_{e},
\end{align}
which is similar to a voltage across a capacitor, we label this the free charge terminal voltage, $V_{e}$. 

Next we consider the magnetic surface current per unit length, which will be apparent at the radial boundary, ($r=r_e$) of the generator. This surface magnetic current will determine the parallel boundary condition, and can be calculated from equation (\ref{emfff}), using the left hand rule to be,
\begin{align}
&\vec{\kappa}_{m}^i=-\frac{\sigma_e^i}{\epsilon_0}\hat{\phi}.
\label{magcurf}
\end{align}

From the integral equations (\ref{eq:V1})$\rightarrow$(\ref{eq:V4}) it is straightforward to derive the boundary conditions. Here subscript ``in" refers to inside the ideal generator and subscript ``out" refers to outside the generator, while the subscript ``$\perp$" refers to the perpendicular components of the field with respect to a surface and the subscript ``$\parallel$" refers to the parallel components of the fields with respect to the surface. We also note that $\vec{E}_{in}=-\frac{\sigma_e^i}{\epsilon_0}\hat{z}$, and that there is no electric surface current (only volume current). Thus, the boundary conditions on the axial surfaces for the ideal voltage source become,
\begin{equation}
E_{out}^{\perp}=0,\label{fcBC1}
\end{equation}
\begin{equation}
\vec{B}_{out}^{\perp}=\vec{B}_{in}^{\perp}=0,\label{fcBC3}
\end{equation}
and the boundary conditions on the radial surface gives,
\begin{equation}
\vec{E}_{out}^{\parallel}=-\vec{\kappa}_{m}^i\times\hat{n}=-\vec{E}_{e}^{i}=-\frac{\sigma_e^i}{\epsilon_0}\hat{z},\label{fcBC4}
\end{equation}
\begin{equation}
\vec{B}_{out}^{\parallel}=\vec{B}_{in}^{\parallel}=0,\label{fcBC2}
\end{equation}

Applying the radial boundary condition, eqn.(\ref{fcBC4}), gives  $\vec{E}_{out}^{\parallel}=-\frac{\sigma_e^i}{\epsilon_0}\hat{z}$. This means the electric field just outside the generator has maximum value on the radial boundary, {\it{despite having infinite conductance}}, similar to fringing in a capacitor. To calculate the magnetic field in the system we can start inside the voltage source and apply the modified Ampere's law given by eqn. (\ref{eq:V2}). Effectively we find that the $\vec{B}$-field caused by the displacement current produced by the time varying $\vec{E}$-field is suppressed by the $\vec{B}$-field caused by the time vary impressed $\vec{E}_e^i$-field (or impressed electrical current), so is small within the electricity generator depending on the aspect ratio. Thus, with no load circuit attached to the generator, the solution is similar to that of a capacitor like dipole, even though we consider an ideal conductivity. In reality the generator will have an internal resistance, so $\vec{E}_T$ will be non-zero, and there will be finite field to match on the axial boundary.

Assuming a harmonic surface charge density of the form, $\sigma_{e}^i=\sigma_{e_0}e^{j\omega_0 t}$, then the terminal voltage may be determined to be,  
\begin{equation}
V_{e}(t)=V_{e_0}e^{j\omega_0 t}=\frac{\sigma_{e_0}d_e}{\epsilon_0}e^{j\omega_0 t}. 
\end{equation}
Likewise, the current inside the voltage source may be determined to be, 
\begin{equation}
I_{e}^i(t)=I_{e_0}e^{j\omega_0 t}=j\omega_0\sigma_{e_0}\pi r_e^2e^{j\omega_0 t}.
\end{equation}
The ratio of the terminal voltage to internal current can be calculated to be $\frac{V_{e}}{I_{e}^i}=\frac{1}{j\omega_0 C_{eff}}$ where $C_{eff}=\frac{\epsilon_0\pi r_e^2}{d_e}$ is similar to a capacitance and $V_e$ lags $I_e^i$ by $\frac{\pi}{2}$. Note, this is not a capacitance, but just the phase relationship between the current and voltage of the generator, an active component where the impressed force maintains the charge separation with no power dissipation, so the internal current and terminal voltage must be out of phase (power factor of $\frac{\pi}{2}$ for no load). Such current and emf generation with no load will create the near field of a Hertzian dipole antenna, which is reactive and exists as stored energy, acting very much like the field of a dynamically charging and discharging capacitor of $C_{eff}$, with the dipole `ends' acting as plates giving a fringing capacitance. To dissipate or radiate power an effective resistive component in the model must be added, which would represent far field radiation or resistive dissipation within the generator. 

\subsection{Short circuit response of the ideal free charge AC generator}

\begin{figure}[t]
\includegraphics[width=1.0\columnwidth]{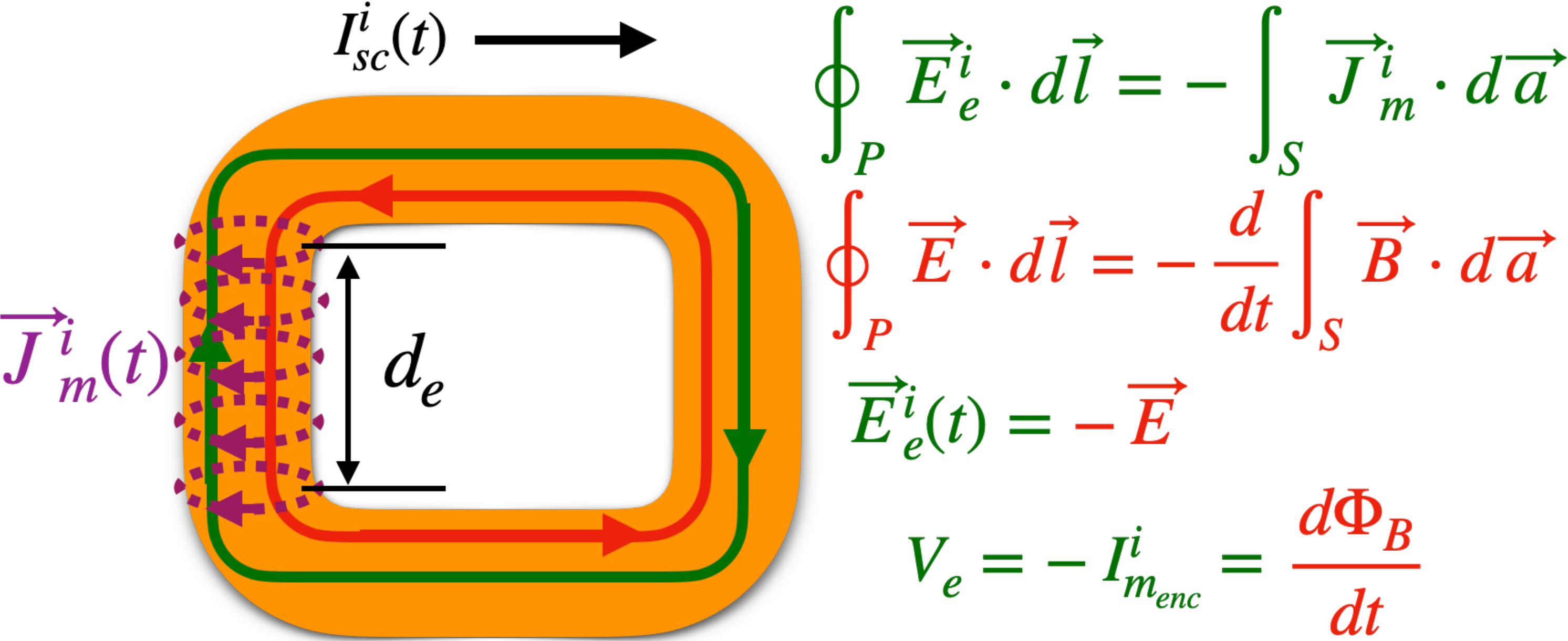}
\caption{Exaggerated short circuit of an AC free-charge voltage generator. Any short circuit will form a loop, which will enclose the impressed magnetic current on the boundary, $\vec{J}_m^i(t)$. Initially, the electrons will flow in the loop driven by, $\vec{E}_e^i$. However, a back emf will be generated, which directly opposes this field, so $\vec{E}_T=\vec{E}_e^i+\vec{E}=0$, and there is no net force on the electrons. The net result is that the voltage supplied by the generator, $V_e$, which exists across the effective inductance, will be determined uniquely by the enclosed magnetic current, given by, $V_e=-I_{m_{enc}}=\frac{d\Phi_B}{dt}$ $=L\frac{dI_{sc}}{dt}+I_{sc}\frac{dL}{dt}$, were $\Phi_B$ is the linked magnetic flux of the loop, and $I_{sc}$ is the short circuit current. The inductance and current flow will be determined by geometry of the loop by implementing Ampere's law.}
\label{SCFC}
\end{figure}

Effectively the calculation in the previous section calculated the open circuit voltage of the AC generator. In reality the generator has a source impedance, which will be designed to have minimal effect, usually represented by a Thevenin or Norton equivalent circuit depending if configured as a voltage or current source. To calculate the Norton equivalent impedance, the short circuit current needs to be calculated and $\vec{J}_f$ in eqn. (\ref{Vss2}) cannot be set to zero and has to be reinstated on the right hand side. For the ideal system, assuming a perfect conductor, the impedance will actually depend on how the voltage terminals are short circuited. Inevitably the short circuit forms a current loop as highlighted in Fig.\ref{SCFC} and thus will have an inductance which will limit the current flow for an AC signal. At DC the finite conductivity of the generator and the metal in the short circuit, will give an effective resistance, which will limit the current flow. Combined with this inductance, the resistance will define the time constant of the short circuit.

If we consider an idealized short circuit for the time-dependent case, with just the equivalent inductance of the loop, then as shown in Fig.\ref{SCFC}, the net force on the electrons will be zero, as a back emf will be produced from a Faraday induced electric field of, $\vec{E}=-\vec{E}_e^i$, so that $\vec{E}_T=0$. Because they have opposing values of curl, Faraday's law becomes, 
\begin{equation}
-\vec{J}_m=\frac{\partial\vec{B}}{\partial t},
\end{equation}
so that the voltage across the loop is given by, 
\begin{equation}
V_e=-I_{m_{enc}}^i=\frac{d\Phi_B}{dt},
\end{equation}
where $\Phi_B$ is the magnetic flux. This is basically the equation for a voltage across the inductor, where 
\begin{equation}
V_e=\frac{d\Phi_B}{dt}=L\frac{dI_{sc}}{dt}+I_{sc}\frac{dL}{dt}. 
\end{equation}
In this case, the free current flow in the loop is determined by the inductance of the loop, calculable from the implementation of Ampere's law with $\vec{J}_f$ of non-zero value.  Even though $\vec{E}_T=0$, the voltage across the short circuit is defined uniquely by the magnetic current boundary source, which best describes the output voltage of an AC generator, rather than the total electric field.

This concept also allows interpretation of Faraday experiments, which either rely on a Coulomb force from the separation of charges opposing the impressed force (open circuit case), or a back emf, from a Faraday force caused by the time rate of change of magnetic flux through a closed loop, which also opposes the impressed force (short circuit case).

\subsection{Example: A cylindrical ideal conductor oscillating in a DC magnetic field}

\begin{figure}[t]
\includegraphics[width=1.0\columnwidth]{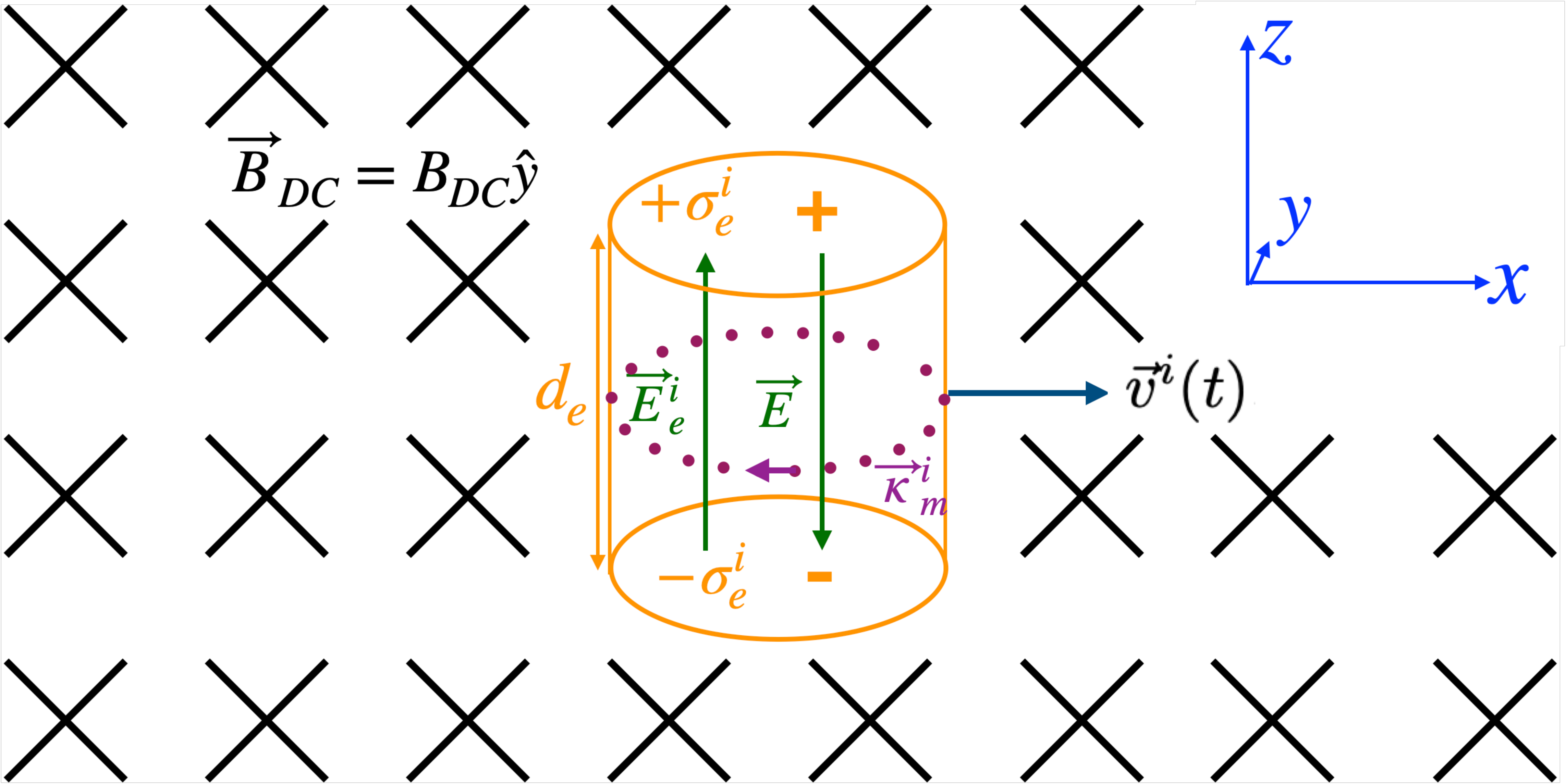}
\caption{A cylindrical ideal conductor (orange) of time-dependent velocity, $\vec{v}^i(t)$, moving in the $\hat{x}$ direction within a DC magnetic field of $\vec{B}_{DC}$ orientated in the $\hat{y}$ direction (black crosses). An emf is generated through the induced Lorentz force per unit charge, $\frac{\vec{F}_e^i}{q}=\vec{E}_e^i=\vec{v}^i(t)\times \vec{B}_{DC}=v_0(t)B_{DC}\hat{z}$ in the $\hat{z}$ direction, which is impressed on the free charges in the conductor. The separation of charges due to the impressed Lorentz force creates a reverse electric field, $\vec{E}$ (green), with an effective magnetic current boundary source, $\vec{\kappa}_m^i$ (magenta).}
\label{LF}
\end{figure}

In this example we use the impressed source technique to analyse the emf induced in a cylindrical ideal conductor due to an external motional kinetic energy or applied force in the quasi-static limit. The conductor generates electricity from its motion due to the impressed Lorentz force, $\vec{F}_e^i$, acting on free charge carriers in the bar as shown in Fig.\ref{LF} and given by,
\begin{equation}
\vec{F}_e^i=q\vec{v}^i(t)\times\vec{B}_{DC}=q\vec{E}_e^i(t). 
\end{equation}
For this case the impressed force per unit charge comes from a driven oscillating mechanical degree of freedom interacting with a DC magnetic field (DC photonic degree of freedom). This in turn creates an oscillating AC photonic degree of freedom characterised by an oscillating emf, $\mathcal{E}(t)$, and given by,
\begin{equation}
\mathcal{E}(t)=E_e^i(t)d_e=d_eB_{DC}v^i(t)
\label{LFemf}
\end{equation}
Assuming simple harmonic motion, $\vec{x}(t)=x_0\sin(\omega_0t)\hat{x}$, then the impressed velocity vector may be written as, 
\begin{equation}
\vec{v}^i(t)=\omega_0x_0\cos(\omega_0 t)\hat{x}. 
\end{equation}
Then given that $\vec{B}_{DC}=B_{DC}\hat{y}$, the impressed Lorentz force per unit charge becomes, 
\begin{equation}
\vec{E}_e^i(t)=\omega_0B_{DC}x_0\cos(\omega_0 t)\hat{z}, 
\end{equation}
creating an emf of,
\begin{equation}
\mathcal{E}(t)=\omega_0B_{DC}d_ex_0\cos(\omega_0 t). 
\end{equation}
The Lorentz force also drives the motion of the charges and hence the impressed internal electrical current within the cylinder of, 
\begin{equation}
\vec{I}_e^i(t)=-\omega_0^2\epsilon_0\pi r_e^2B_{DC}x_0\sin(\omega_0 t)\hat{z}. 
\end{equation}
Note the current lags the voltage by $\frac{\pi}{2}$ as expected. The oscillating separation of $\pm$ charges creates an electric field, which opposes the impressed force so the net force on the oscillating charges is zero, which means there is no work done in driving the internal electrical current. This means that there is only reactive power in the electrical degree of freedom as there is no load to dissipate active power. However, as a consequence of this oscillating current in the DC magnetic field, there is an additional Lorentz force acting on the mass of the cylinder, $\vec{F}_B(t)$, which is given by,
\begin{equation}
\vec{F}_B(t)=d_e\vec{I}_e^i(t)\times\vec{B}_{DC},
\label{LB}
\end{equation}
due to the mechanical motion. This force is in the same direction, but out of phase with the velocity, given by, 
\begin{equation}
\vec{F}_B(t)=\omega_0^2\epsilon_0\mathcal{V}_eB_{DC}^2x_0\sin(\omega_0 t)\hat{x}, 
\end{equation}
where $\mathcal{V}_e$ is the volume of the conductor. In the ideal case there is no work done unless a load is attached to the generator. Of course the analysis ignores any real material effects, such as finite conductivity, skin effect, kinetic inductance etc., which occur at high frequencies. 

Thus, in the real case the oscillating conductor has an effective impedance, including a source resistance due to DC conductivity, which will dominate at low frequencies, this will be a source of damping as a component of, $\vec{F}_B(t)$ would be in phase with the velocity, which also means a component of the current and voltage would be in phase. At higher frequencies the resistance will increase due to the skin effect, and the cylinder will have an inductance. It is not our intention to analyse these effects in this paper, which will contribute to the source impedance of a real voltage source or generator. As discussed previously, the cylindrical generator will also emit an $\vec{E}(t)$-field outside, which will be in the form of the near field of an oscillating Hertzian dipole.

\section{Generation of Electricity from Bound Charge}

\subsection{The DC Electret}

\begin{figure}[t]
\includegraphics[width=1.0\columnwidth]{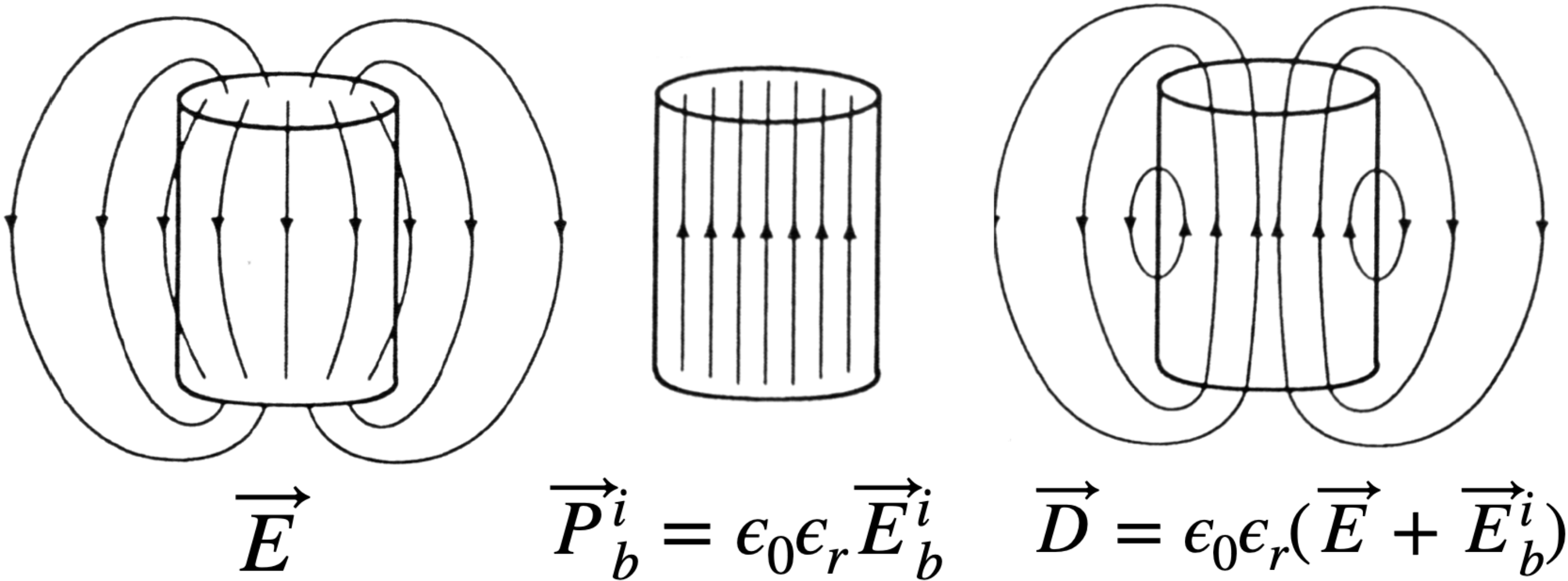}
\caption{From left to right, 3D sketch of the  $\vec{E}$, $\vec{P}_{_b}^i$ and $\vec{D}$ fields inside and outside a cylindrical bar electret (reproduced from the solution manual of  \cite{GriffBook}). Assuming the impressed polarization, $\vec{P}_{_b}^i$, is constant within the electret and along the cylindrical $z-$axis of the bar in the positive direction. Note, $\vec{P}_{_b}^i$ is only defined inside the electret.}
\label{elecfield2}
\end{figure}

A DC bound charge voltage source is essentially a bar electret. The ideal bar electret exhibits a permanent electrical dipolar field as shown in Fig.\ref{elecfield2}, due to an impressed macroscopic polarization, $P_b^i$, and has overall charge neutrality\cite{Electrets}. Some common ways to impress a polarization and make an electret, is to heat a polar dielectric material under the influence of a large electric field (thermo electret) \cite{Jefimenko1980} or through piezoelectricity (piezo electret) \cite{Sessler2016}. For the former, once cooled and removed from the electric field a net polarization will be maintained, while the later is maintained through the application of strain. Other forms of electrets include the magneto-electret and the magneto active electret \cite{Monkman_2017} as well as ferroelectric electrets \cite{Asanuma2013,Graz2016,C6TA09590A}. The electret thus becomes a bound charge voltage source and is useful for supplying DC bias reducing the requirement for high external DC voltages \cite{mi11030267,Jean-Mistral2012}, and can supply a current and be discharged in a similar way to a battery \cite{Gross62}. In fact batteries with solid electrolytes are not too dissimilar to a DC electret. For example, when a ferroelectric material becomes permanently polarized, it undergoes a phase change, which corresponds to the crystal structure breaking a certain symmetry under phase transition \cite{Zhao2020}. Likewise a rechargeable battery with a solid electrolyte undergoes induced symmetry breaking of the electrolyte structure, such as polyhedron distortions upon charge and discharge \cite{LIU2020}. Some solid batteries even have electrolyte of perovskite structure \cite{Xu18815,Ahmad:2018ro}, similar to the structure of common ferroelectric materials.

The bar electret as shown in Fig.\ref{elecfield2} is the electrostatic analogue of the bar magnet. Here we assume a lossless dielectric media so ideally there is no free charge or current in the system and hence they are set to zero, in this case Maxwell's equations, (\ref{V1})$\rightarrow$(\ref{Aux}), become
\begin{align}
&\vec{\nabla}\cdot\vec{D}=0,\label{Ee1}\\
&\vec{\nabla} \times \vec{E}=0, \label{Ee2}
\end{align}
where
\begin{align}
&\vec{D}=\epsilon_0\vec{E}+\epsilon_0\chi_e\vec{E}+\vec{P}_b^i=\epsilon_0\epsilon_r\vec{E}+\vec{P}_b^i, 
\label{Ee3}
\end{align}
for a linear dielectric material with a permanent polarization, $\vec{P}_b^i$, which is independent of the electric field, $\vec{E}$.

\subsubsection{Impressing a Source Term into static Maxwell's Equations to Describe an Electret}

Equations (\ref{Ee1})$\rightarrow$(\ref{Ee3}) seem incomplete, as there is a lack of a source term on the right hand side of the equations. For the analogue bar magnet, Maxwell's equations give $\vec{\nabla}\cdot\vec{B}=0$, $\vec{\nabla}\times\vec{H}=0$ and $\vec{\nabla}\times\vec{B}=\mu_0\mu_r\vec{\nabla}\times\vec{M}_S$ for a linear magnetic material with a permanent magnetization $\vec{M}_S$. The vector $\vec{M}_S$ is in actual fact an impressed source as energy needs to be added by an external force to permanently magnetize the material, and we may identify an impressed bound current, $\vec{J}_b^i=\vec{\nabla}\times\vec{M}_S$ so that Ampere's law becomes $\vec{\nabla}\times\vec{B}=\mu_0\mu_r\vec{J}_b^i$ where the effective impressed bound current at the radial boundary of the bar magnet sources the magnetic field. Thus, the fields in the bar magnet could also be represented in Fig.\ref{elecfield2} with the following substitutions $\vec{B}\rightarrow\vec{D}$, $\vec{H}\rightarrow\vec{E}$ and $\vec{M}_S\rightarrow\vec{P}_b^i$. Now we can identify how static Maxwell's equations need to be generalised to be able to describe a system with a permanent polarization as an impressed source, that is to take the curl of eqn. (\ref{Ee3}) and combine with eqn. (\ref{Ee2}) to give $\vec{\nabla}\times\vec{D}=\vec{\nabla}\times\vec{P}_b^i$. One can identify an impressed magnetic bound (subscript $mb$) current, $\vec{J}_{mb}$, boundary source at the radial boundary of the electret, from,
\begin{align}
\vec{\nabla}\times\vec{P}_b^i=-\epsilon_0\epsilon_r\vec{J}_{mb}^i
\label{modfara}
\end{align}
so that a modified Faraday's law becomes, 
\begin{align}
\vec{\nabla}\times\vec{D}=-\epsilon_0\epsilon_r\vec{J}_{mb}^i~~\text{(left hand rule)}.
\label{modfar}
\end{align}

There is no free charge in this system so the divergence of $\vec{D}$ is zero. However, the divergence of $\vec{E}$ will be non zero due to the separation of bound charge, which has in general two components, $\rho_{b\chi_e}$, due to the dielectric susceptibility of the material and, $\rho_{b_{P^i}}$, the bound charge driven by the AC impressed polarization vector, $\vec{P}^i_e(t)$, so that,
\begin{equation}
\epsilon_0\vec{\nabla}\cdot\vec{E}=\rho_{b\chi_e}+\rho_{b_{P^i}},
\end{equation}
and
\begin{equation}
\epsilon_0\epsilon_r\vec{\nabla}\cdot\vec{E}=-\vec{\nabla}\cdot\vec{P}_b^i=\rho_{b_{P^i}}.
\end{equation}

In a similar way to the free charge voltage source, the forces in the bound charge system may be defined using equation (\ref{freeE2}) \cite{GriffBook,RHbook2012}. Thus, the total force per unit charge, $\vec{E}_T$ acting on the bound charges is given by,
\begin{equation}
\vec{E}_T=\vec{E}+\vec{E}_b^i,
\label{boundE2}
\end{equation}
Accordingly the impressed electric field acting on bound charge, $\vec{E}_b^i$, may be identified to be related to the permanent source polarization $\vec{P}_b^i$ by,
\begin{align}
\vec{E}_b^i=\vec{P}_b^i/(\epsilon_0\epsilon_r),
\label{forceb}
\end{align}
and then if we multiply equation (\ref{boundE2}) through by the permittivity, $\epsilon_0\epsilon_r$, we obtain exactly eqn. (\ref{Ee3}) (where $\vec{D}=\epsilon_0\epsilon_r\vec{E}_T$), the well known constitutive relation between the $\vec{D}$-field, $\vec{E}$-field and $\vec{P}$-field in the electret. Thus, equation (\ref{boundE2}), which balances the forces in the voltage source is essentially on the same footing as a constitutive relationship between fields given by eqn. (\ref{Ee3}). In this case the modified static Faraday's law may be written as,
\begin{align}
\vec{\nabla}\times\vec{E}_T=-\vec{J}_{mb}^i,
\label{frdlawmod}
\end{align}
which is equivalent to eqn. (\ref{modfar}). Thus as emphasized in the appendix, the DC voltage source has two components of force per unit charge in the system. The impressed external force per unit charge, $\vec{E}_b^i$, with an electric vector potential, which generates an emf, and applies the force to seperate the $\pm$ charges, and the electric field, $\vec{E}$, which is sourced by the separated $\pm$ bound charges, which exhibits a scalar electric potential.

Revisiting the magnetic analogue of the DC electret (a permanent magnet), where, $\vec{\nabla}\times\vec{B}=\mu_0\mu_r\vec{J}_b^i$, introduces a bound (or Amperian) current \cite{Birch_1985}. In this example the bound current is an effective impressed electrical current, and there is no real current flow, especially this is highlighted by the fact that a permanent magnet is usually not a good conductor. The collective alignment of spins, which creates this bound current occurs due to an impressed force magnetising the material (in this case a magnetomotive force). This bound current is treated exactly like a real current when analysing macroscopic electrodynamic equations, like in magnetic circuit analysis. For example, the bound electrical current in a bar magnet acts as a source term for the fields of the macroscopic magnetic dipole. In a similar way, the magnetic current defined and introduced in eqns. (\ref{modfara}), (\ref{modfar}) and (\ref{frdlawmod}) is not a real magnetic current, and of course it cannot be as monopoles as far as we know do not exist. This term is an effective magnetic current that defines the boundary conditions, and in analogy can be used when analysing macroscopic electrodynamic equations, like in electric circuit analysis. Similarly, the bound magnetic current in a bar electret acts as a source term for the fields of the macroscopic electric dipole. Another interesting point is that the free charge voltage source discussed previously can be also thought of as a macroscopic dipole, with a free charge polarization, with a similar magnetic current boundary source.

To calculate the fields in the DC electret, a numerical calculation is necessary, similar to that of a cylindrical bar magnet. However, to calculate the circuit properties of the voltage source one just need to implement the integral form of equation (\ref{frdlawmod}), which gives $\mathcal{E}=-I^i_{mb}=\frac{1}{\epsilon_0\epsilon_r}\oint_P\vec{P}_b^i\cdot d\vec{l}$ independent of the electric field since $\oint_P\vec{E}\cdot d\vec{l}=0$. The other parameter that is necessary to calculate is the source impedance, which will just be the capacitance of the electret with the source polarization removed. To first order this capacitance may be calculated assuming a constant $\vec{E}$ field within the dielectric, which ignores fringing, and assumes an aspect ratio of a thin polarized plate. Nevertheless, the calculation uncertainties in this case is with the source impedance and not the emf. However, in a general system one might expect a numeric calculation is necessary if the polarization and electric fields cannot be assumed constant, similar to a bar magnet.

\subsection{AC Generation of Electricity with Time Varying Bound Charge}

Next, we generalise the concept of the electret to a time varying impressed permanent polarization, $\vec{P}_b^i(t)$, along the lines to what was undertaken in the work of Zhong Lin Wang \cite{WANG201774}. This system describes an AC electricity generator based on oscillating bound charge. The starting point is to consider Maxwell's equations for a linear isotropic dielectric media with a time varying permanent polarization and an impressed magnetic current boundary source, in a similar way to the static version discussed previously (also see appendix). First we consider the contributions to Gauss' Law for the three fields, $\vec{D}(t)$, $\vec{E}(t)$ and $\vec{P}_b^i(t)$, which is modified in the same way as the DC case,
\begin{align}
&\vec{\nabla}\cdot\vec{D}=0,~~\vec{\nabla}\cdot\vec{E}=\frac{\rho_{b_{P^i}}}{\epsilon_0\epsilon_r}~~\text{and}~\vec{\nabla}\cdot\vec{P}_b^i=-\rho_{b_{P^i}}. \label{DM7}
\end{align}
Next, Ampere's law is modified along the lines as first suggested by Wang \cite{WANG201774},
\begin{align}
&\vec{\nabla} \times \vec{B}-\mu_0\epsilon_0\epsilon_r\frac{\partial\vec{E}}{\partial t}-\mu_0\frac{\partial\vec{P}_b^i}{\partial t}=0.\label{DM8}
\end{align}
The  magnetic Gauss' law remains unmodified, while Faraday's law is modified so,
\begin{align}
&\frac{1}{\epsilon_0\epsilon_r}\vec{\nabla} \times \vec{D}+\frac{\partial \vec{B}}{\partial t}=-\vec{J}_{mb}^i, \label{DM10}
\end{align}
with
\begin{align}
&\vec{\nabla} \times \vec{E}+\frac{\partial \vec{B}}{\partial t}=0~\text{and}~\vec{\nabla} \times \vec{P}_b^i=-\epsilon_0\epsilon_r\vec{J}_{mb}^i. \label{DM10a}
\end{align}
The last term in equation (\ref{DM10}) is the impressed magnetic current, which acts as a source of the system, to create an AC voltage output. The modified Maxwell's equations with impressed sources may also be written in terms of the total force per unit charge, $\vec{E}_T=\vec{E}+\vec{E}_b^i$, for the time-dependent electret (also see the appendix) as,
\begin{align}
&\vec{\nabla}\cdot\vec{E}_T=0\\
&\vec{\nabla} \times \vec{B}-\mu_0\epsilon_0\epsilon_r\frac{\partial\vec{E}_T}{\partial t}=0\\
&\vec{\nabla}\cdot\vec{B}=0\\
&\vec{\nabla} \times \vec{E}_T+\frac{\partial \vec{B}}{\partial t}=-\vec{J}_{mb}^i,
\label{magC}
\end{align}
Here the $\frac{\partial \vec{B}}{\partial t}$ term in eqn.(\ref{DM10}) and (\ref{magC}) can be identified as the magnetic displacement current. In this system, the effective magnetic current source term, $\vec{J}_{mb}^i$ exists on the radial boundary of the electret, and drives the impressed electric filed, $\vec{E}^i_b$ by the left hand rule and also sets the boundary condition for the parallel components of the fields on the radial boundary. 

\subsection{Boundary Conditions}

The boundary conditions of the fields on the normal and parallel surfaces of the electret can be calculated from the integral version of equations (\ref{DM7})$\rightarrow$(\ref{DM10a}), which are given by,
\begin{equation}
\oiint_S\vec{D}\cdot d\vec{a} = 0,
\end{equation}
\begin{equation}
\oiint_S\vec{E}\cdot d\vec{a} = \frac{Q^i_{b_{P^i}}}{\epsilon_0\epsilon_r}, ~~
\oiint_S\vec{P}_b^i\cdot d\vec{a} = -Q^i_{b_{P^i}}\label{Intel1},
\end{equation}
\begin{equation}
\oint_P \vec{B}\cdot d\vec{l}-\mu_0\frac{d}{dt}\int_S\vec{D}\cdot d\vec{a}=0\label{Intel2}
\end{equation}
\begin{equation}
\oiint_S \vec{B}\cdot d\vec{a} = 0\label{Intel3},
\end{equation}
\begin{equation}
\frac{1}{\epsilon_0\epsilon_r}\oint_P\vec{D}\cdot d\vec{l}+\frac{d}{dt}\int_S\vec{B} \cdot d\vec{a}=-I^i_{mb},\label{Intel4a}
\end{equation}
\begin{equation}
\oint_P\vec{E}\cdot d\vec{l}+\frac{d}{dt}\int_S\vec{B} \cdot d\vec{a}=0, ~~
\frac{1}{\epsilon_0\epsilon_r}\oint_P\vec{P}_b^i\cdot d\vec{l}=-I^i_{mb}
\end{equation}
From these integral equations it is straightforward to derive the modified boundary conditions given in (\ref{elBC1})$\rightarrow$(\ref{eleBC5}). Here subscript ``in" refers to inside the bar electret and subscript ``out" refers to outside the bar electret.,

\begin{equation}
D_{out}^{\perp}=D_{in}^{\perp},
\label{elBC1}
\end{equation}
\begin{equation}
\epsilon_{out}E_{out}^{\perp}-\epsilon_{in}E_{in}^{\perp}=\sigma_{b_{P^i}},~~\epsilon_{in}E_{b_{in}}^{i\perp}=P_{b_{in}}^{i\perp}=\sigma_{b_{P^i}}
\label{elBC1a}
\end{equation}
\begin{equation}
\vec{B}_{out}^{\parallel}=\vec{B}_{in}^{\parallel},\label{elBC4}
\end{equation}
\begin{equation}
B_{out}^{\perp}=B_{in}^{\perp}.\label{elBC2}
\end{equation}
\begin{equation}
\vec{D}_{{out}}^{\parallel}-\vec{D}_{{in}}^{\parallel}=-\epsilon_0\epsilon_r\vec{\kappa}_{mb}^i\times\hat{n},\label{eleBC3}
\end{equation}
\begin{equation}
\vec{E}_{{out}}^{\parallel}=\vec{E}_{{in}}^{\parallel},~~\vec{E}_{b_{in}}^{i\parallel}=\frac{\vec{P}_{b_{in}}^{\parallel}}{\epsilon_{in}}=-\vec{\kappa}_{mb}^i\times\hat{n},\label{eleBC5}
\end{equation}

In the following we show how the values of the electromagnetic fields, output voltage and magnetic current may be calculated with the aid of the constitutive relations and boundary conditions that define an ideal electret.  

\subsubsection{Axial Boundaries}

\begin{figure}[t]
\includegraphics[width=1.0\columnwidth]{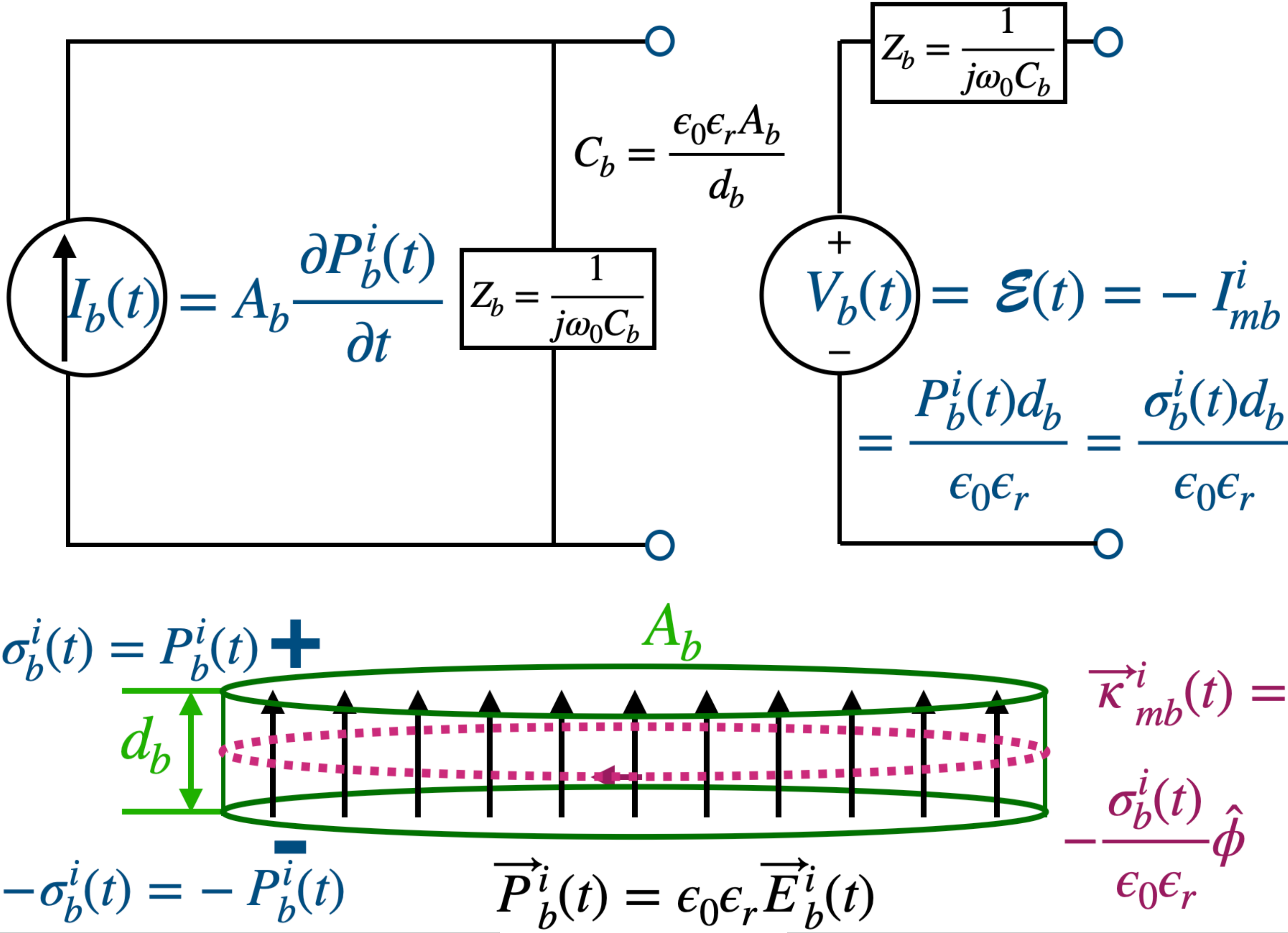}
\caption{Top left and right show respectively the Norton and Thevenin equivalent circuits, for the ideal AC bar electret of cross sectional area $A_b$ and length $d_b$ shown underneath. The associated impressed force per unit charge $\vec{E}_{b}^i(t)$ supplies the force to seperate the bound charges $\sigma^i_b$ resulting in an electret with a time varying permanent polarisation of $\vec{P}_b^i(t)={P}_b^i(t)\hat{z}$, which can be modelled as an impressed magnetic current boundary source, $\vec{\kappa}_{mb}^i(t)$. The Thevenin equivalent circuit is pictorially shown with the open circuit voltage, $V_b(t)=\mathcal{E}(t)=-I_{mb}^i$, and effective source impedance, $Z_{b}$, which is calculated in the text and is equivalent to the capacitance of the electret without a permanent polarization.}
\label{ele}
\end{figure}

Assuming the impressed polarization is constant of the form $\vec{P}_b^i={P}_b^i\hat{z}=\epsilon_0\epsilon_rE_{b}^{i}\hat{z}$, at the top and bottom axial boundaries, as shown in Fig.\ref{ele}, we can use eqn.(\ref{elBC1}) to obtain a relationship between $\vec{E}$ and $\vec{E}_{b}^{i}$, where
\begin{equation}
\epsilon_0\vec{E}_{out}\cdot\hat{n}=\epsilon_0\epsilon_r(\vec{E}_{in}+\vec{E}_{b}^i)\cdot\hat{n}
\label{axBC}
\end{equation}
To take the analysis a step further one really needs a numeric solution for the general case, as another equation is needed to solve the problem. In general, $\vec{E}_{out}$ is non zero as highlighted in Fig.\ref{elecfield2}, and $\vec{E}_{in}$ points in the opposite direction to $\vec{E}_{out}$ and $\vec{E}_{b}^i$ and in general  $|\vec{E}_{b}^i|\ge|\vec{E}_{in}|$. In our example we have assumed a constant $\vec{P}_b^i$ along the $z$-axis, this is actually an approximation for a thin plate, where we ignore fringing, which means $\vec{E}_{out}\approx0$ so that $\vec{E}_{in}\approx-\vec{E}_{b}^i$. This limit gives as a simple way of calculating the Thevenin equivalent circuit for an electret.

Matching this condition gives the following relation between vectors at the axial boundary,
\begin{equation}
\vec{E}_{{out}}=0~~\text{and}~~\vec{E}_{{in}}=-\vec{E}^i_{b}.
\label{Erat}
\end{equation}
so that $\vec{D}=0$ above and below the axial boundaries. To calculate the impressed bound surface current, $\pm\sigma^i_b$ at the upper and lower axial end faces respectively, we use
\begin{equation}
\sigma^i_b=\vec{P}^i_{b}\cdot\hat{n}=\epsilon_0\epsilon_rE^i_b,
\label{PT}
\end{equation}
as $\hat{n}=\hat{z}$, while on the lower axial boundary the value is negative as $\hat{n}=-\hat{z}$.

The polarization current density may be calculated from the time rate of change of eqn. (\ref{PT}) to be, $\vec{J}_P=\frac{\partial \vec{P}_b^i}{\partial t}=\epsilon_0\epsilon_r\frac{\partial E_b^i}{\partial t}\hat{z}$, and if it has a cross sectional area of $A_b$ the effective polarization current through the voltage source is given by,
\begin{equation}
I_{P}=A_b\frac{\partial P^i_{b}}{\partial t}=A_b\frac{\partial \sigma^i_{b}}{\partial t}=A_b\epsilon_0\epsilon_r\frac{\partial E_b^i}{\partial t}
\label{Id}
\end{equation}

\subsubsection{Radial Boundaries}

On the radial boundary we can determine that the impressed surface magnetic current density from equation (\ref{eleBC5}) to be,
\begin{equation}
\vec{\kappa}_{m}^i=-E_{b_{in}}^{i}\hat{\phi}=-\frac{\sigma^i_{b}}{\epsilon_0\epsilon_r}\hat{\phi},\label{Km}
\end{equation}
and the electric fields, $\vec{E}$ to be,
\begin{align}
\vec{E}_{in}=\vec{E}_{out}=-E_b^i\hat{z}=-\frac{\sigma^i_{b}}{\epsilon_0\epsilon_r}\hat{z}
\label{Ebrad}
\end{align}
This is of similar form to the free charge voltage source, with $\sigma^i_e\equiv\frac{\sigma_b^i}{\epsilon_r}$, because both have $\vec{E}_T=0$ above and below the axial boundary. This also means that the total displacement current, $\vec{J}_D=0$, above and below the axial boundary and hence there will be no $\vec{B}$ field within the voltage source. 

\subsubsection{Equivalent Circuit}

To calculate the emf we need to integrate around the radial boundary, with $\frac{\partial \vec{B}}{\partial t}=0$ and therefore the emf may be calculated from,
\begin{align}
&\mathcal{E}=-I^i_{mb}=\oint_P\vec{E}^i_{b}\cdot d\vec{l},
\label{emfbb}
\end{align}
where $I^i_{mb}= \int_S\vec{J}_{m}^i\cdot d\vec{a}$, is the enclosed magnetic current. Defining the length of the voltage source as $d_b$, the induced emf is given by,
\begin{align}
&\mathcal{E}=E_b^i d_b=\frac{\sigma^i_{b}d_b}{\epsilon_0\epsilon_r}
\label{emfb}
\end{align}
Note, $\vec{J}_D=0$ is an approximation, and the value will depend on the aspect ratio, thus for any finite electret there will be in fact a finite $\vec{D}$ field (and hence finite $\vec{E}_T$ field) above the electret, as shown in Fig.\ref{elecfield2}. For these cases a small magnetic field will be generated within the electret due to the time dependence.

Now assuming the charge density oscillates harmonically such that $\sigma^i_b= \sigma_{b0}e^{j\omega_0}$, we can calculate the Thevenin equivalent voltage, where the open circuit voltage across the electret is equivalent to the emf calculated in eqn. (\ref{emfb}) to give,
\begin{align}
V_{b}=\frac{\sigma_{b0}d_b}{\epsilon_0\epsilon_r}e^{j\omega_0 t}
\end{align}
Also by short circuiting the electret, a free charge will oscillate in the short circuit wire equivalent to the polarization current, as the $\vec{E}_{in}$ field is shorted to zero, the Norton equivalent current source can be calculated to be,
\begin{align}
I_{b}=A_b\frac{\partial \sigma^i_{b}}{\partial t}=j\omega_0A_be^{j\omega_0 t}.
\end{align}
Ignoring any small resistive or inductive effects, the source impedance may be calculated to be,
\begin{align}
Z_{b}=\frac{V_b}{I_b}=\frac{1}{j\omega_0C_{b}} \ \ \ C_{b}=\frac{\epsilon_0\epsilon_rA_b}{d_b}.
\label{ZBS}
\end{align}
Note, the impedance can also be calculated by setting the voltage source to zero (or permanent polarization, $P_b^i=0$), and calculating the capacitance of the left over dielectric, which is equivalent to eqn. (\ref{ZBS}). This system is a Hertzian dipole with similar characteristics to the free charge system.

\subsection{Example: A piezoelectric nano generator (PENG)}

\begin{figure}[t]
\includegraphics[width=0.7\columnwidth]{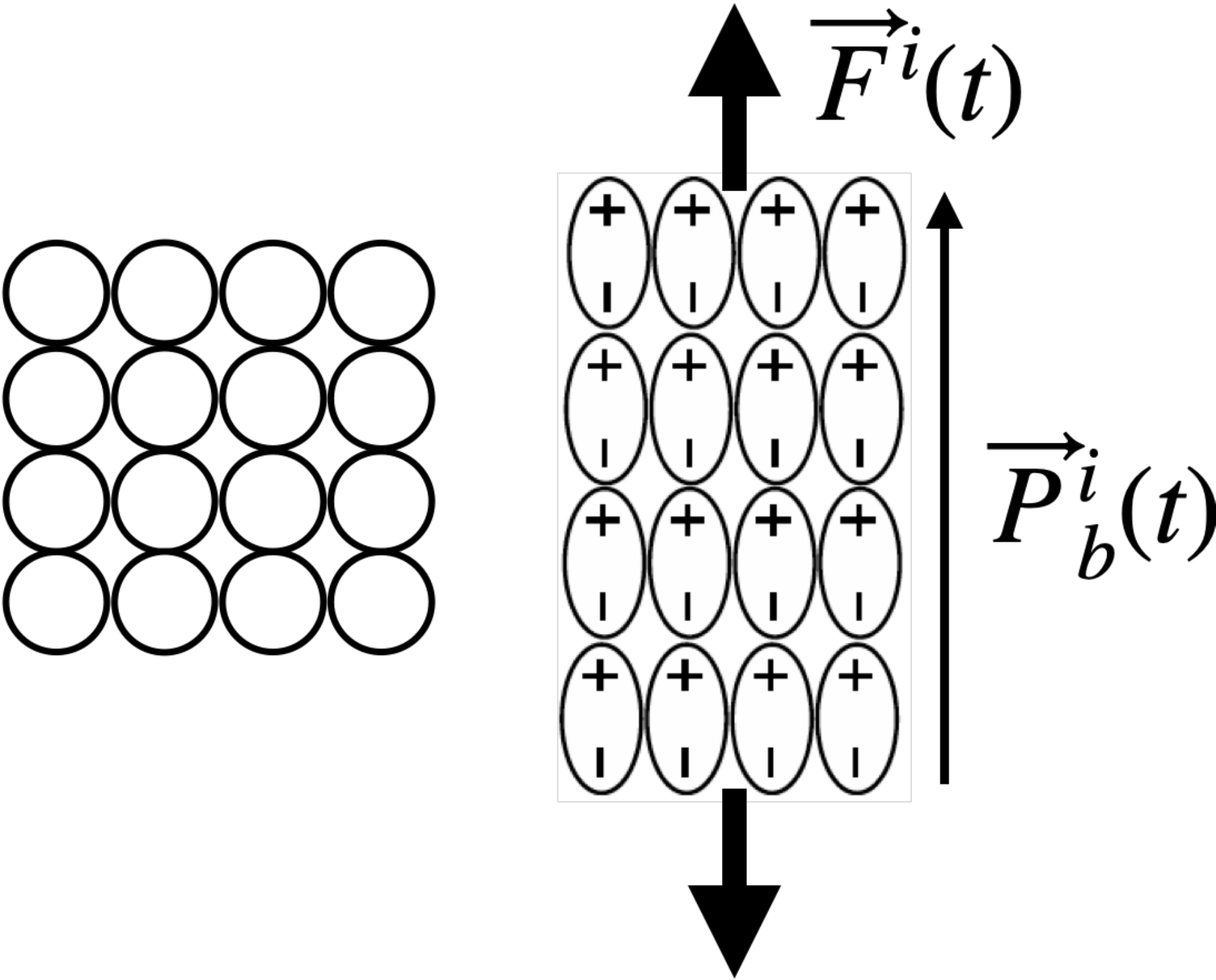}
\caption{Exaggerated diagram of an impressed mechanical load, $\vec{F}^i(t)$, distorting a cubic structure (left) to induce an impressed polarization, $\vec{P}_b^i(t)$ (right), which is the principle of PENG.}
\label{piezo}
\end{figure}

One way to generate the time varying electret as discussed is through piezoelectricity, which has become an important way to undertake energy harvesting \cite{Ertuk2011}. In this example we consider a direct piezoelectric effect where polarization in certain materials can also be induced by mechanical loads as shown in Fig.\ref{piezo}. An external impressed load (force, stress or strain) if time harmonic, will cause a time harmonic variation of impressed polarization, $\vec{P}_b^i(t)$. Thus, the piezoelectric effect is explained through the coupling of electric and elastic phenomena, and thus in general is a tensorial theory combining continuum mechanics and electrodynamics leading to many complex effects beyond the scope of our discussion, see standard text books such as \cite{Ertuk2011,ikeda1996,Yang2018}. For the purposes of this work, we assume a basic mathematical formulation\cite{Dahiya2013}, where dimensional effects are small and the process is linear and in one dimension along the $z$-axis. For this ideal example the constitutive relations are,
\begin{align}
P_{b_{z}}^i=d_{pe}\times T=d_{pe}\times c_{pe} \times S=e_{pe} \times S \label{pe1}\\
D_z=P_{b_{z}}^i+\epsilon_0\epsilon_rE_z \label{pe2}.
\end{align}
Here, $d_{pe}$ is the piezoelectric strain coefficient, $T$ is the stress to which piezoelectric material is subjected, $c_{pe}$ is the elastic constant relating the generated stress, $T$, and the applied strain, $S$, ($T = c_{pe}\times S$) $s_{pe}$ is the compliance coefficient, which relates the deformation produced by the application of a stress ($S = s_{pe}\times T$), and $e_{pe}$ is the piezoelectric stress constant. In this case equation (\ref{pe2}) is of the same form as eqn. (\ref{Ee3}), where the magnitude and time dependence of the electret polarization is dependent on the external mechanical load.

A general theory for potential and fields has been developed for PENG in \cite{WANG201774,WANG2020104272}, which focuses on the generation of the displacement current, $\vec{J}_D=\frac{\partial D_z}{\partial t}=\epsilon_0\epsilon_r\frac{\partial E_z}{\partial t}+\frac{\partial P^i_{b_z}}{\partial t}$. However, in the no-load situation, with a flat aspect ratio, we have shown in the last section that $\vec{J}_D\approx0$, with the open circuit voltage, $V_{oc}=\mathcal{E}=\frac{P_{b_{z}}^i d_b}{\epsilon_0\epsilon_r}=\frac{\sigma^i_{b}d_b}{\epsilon_0\epsilon_r}$, calculable from the modified Faraday's law, given by equations (\ref{emfbb}) and (\ref{emfb}), which is consistent with that derived in \cite{WANG201774}. This calculation highlights, when the terminals are short circuited, the free current that flows will be equal to the polarization current, $I_P=A_b\frac{\partial P^i_{b_z}}{\partial t}$, driven by the induced emf, $\mathcal{E}(t)$ as indicated by the Norton and Thevenin equivalent circuits shown in Fig.\ref{ele}. Note it is generally assumed that the external strains do not significantly perturb the dimensions\cite{WANG201774}, so to fist order the source impedance shown in Fig.\ref{ele} is constant. Also, any inductive or resistive short circuit properties will be small compared to the source capacitance.

Typical parameters for an AC electret based on vibrational energy include; 1) Based on charged resonantly vibrating cantilevers with vibrations of $0.1g (1m/s^2)$, can harvest up to $30\mu W$ per gram of mobile mass, with a source capacitance between $1-8 pF$\cite{Boisseau_2011}; 2) A vibration-driven polymer energy harvester\cite{Suzuki2015,Kashiwagi_2011}, obtained output power as large as $100\mu W$ at $30Hz$ and $0.15 g$ acceleration.  3) The flexible triboelectric generator\cite{FAN2012328} has attained an electrical output peak voltage of $3.3V$ and current of $0.6 mA$ with a peak power density of $10.4 mW/cm^3$. For such devices, output impedances are capacitive and typically of the order of 100 $M\Omega$.

\section{Conclusion}

We have explored the electrodynamics of bound and free charge electricity generators and voltage sources. The external input to the system was represented by an impressed force per unit charge, which converts the external energy into electromagnetic energy and may be considered as a non-conservative electric field vector, or emf per unit length, with an electric vector potential. The source term is necessarily impressed into Maxwell's equations as an effective magnetic current boundary source, which sources the resulting charge distribution and emf produced by the generator, resulting in a modification of Faraday's law, the constitutive relations and hence Maxwell's equations. 

\section*{Acknowledgements}

This work was funded by the Australian Research Council Centre of Excellence for Engineered Quantum Systems, CE170100009 and  Centre of Excellence for Dark Matter Particle Physics, CE200100008. We also thank Professor David Griffiths for allowing the reproduction of his figures and we thank Professor Ian McArthur for his analysis and comments on the manuscript. 

\section*{References}

\section{Appendix A: Two Potential Formulation}

The general two potential formulation of impressed current and fields has been discussed in detail in standard text books on Electrical Engineering \cite{ECJ,RHbook2012,Balanis2012,Kudryavtsev2013}. The two potential formulation is used in electrodynamics to model electricity generation in circuit and antenna theory, when there is conversion of external energy into electromagnetic energy through non conservative processes as discussed in the main body of this paper. It has also been used to describe duality in electrodynamics and axion electrodynamics\cite{Cabibbo1962,Keller2018,TobarModAx19,Asker2018}, and was recently applied to electricity generation \cite{Tobar2021}.

Using superposition, we can consider the electric and magnetic current sources separately. So setting the magnetic sources to zero, the electric and magnetic fields may be written in terms of the magnetic vector potential, $\vec{A}$, and the electric scalar potential, $\phi$,
\begin{equation}
\begin{array}{l}
{\vec{E}_A=-\nabla \phi-\frac{\partial\vec{A}}{\partial t}} \\
{\vec{B}_A=\nabla \times \vec{A}}.
\label{EBA}
\end{array}
\end{equation}
Then by setting the electric sources to zero the electric and magnetic fields may be written in terms of the electric vector potential, $\vec{C}$, and the magnetic scalar potential, $\phi_m$,
\begin{equation}
\begin{array}{l}
{\vec{E}_{C}=-\frac{1}{\epsilon_0}\nabla \times \vec{C}} \\
{\vec{B}_{C}=-\mu_0\nabla\phi_m-\mu_0\frac{\partial\vec{C}}{\partial t}}
\label{EBAtilde}
\end{array}
\end{equation}

The total electric and magnetic fields may be calculated using the principle of superposition and are given by \cite{RHbook2012,Balanis2012};
\begin{equation}
\vec{E}_{Tot}=\vec{E}_A+\vec{E}_{C}=-\nabla \phi-\frac{\partial\vec{A}}{\partial t}-\frac{1}{\epsilon_0}\nabla\times\vec{C}
\label{ETpot}
\end{equation}
\begin{equation}
\vec{B}_{Tot}=\vec{B}_A+\vec{B}_C=-\mu_0\nabla\phi_m-\mu_0\frac{\partial\vec{C}}{\partial t}+\nabla\times\vec{A}.
\label{BTpot}
\end{equation}

Considering the electric field given by equation (\ref{ETpot}), in the quasi-static limit we can ignore the time-dependent terms and the main source terms are due to the charge distributions defined by the electric charge and the effective magnetic current, with the electric vector potential given by \cite{RHbook2012,Balanis2012},
\begin{equation}
\vec{C}( \vec{r},t )=\frac {\epsilon _ {0}} {4\pi} \int _ {\Omega } \frac{ \vec{J}_{m}^i \left( \vec{ r } ^ { \prime } , t ^ { \prime } \right) } { \left| \vec{ r }-\vec{ r } ^ {\prime} \right| } \mathrm{d}^{3} \vec{r}^{\prime}.
\end{equation}
and the electric scalar potential given by,
\begin{equation}
\phi( \vec{r},t )=\frac {1} {4\pi\epsilon _ {0}} \int _ {\Omega } \frac{ \rho\left( \vec{ r } ^ { \prime } , t ^ { \prime } \right) } { \left| \vec{ r }-\vec{r} ^ {\prime} \right| } \mathrm{d}^{3} \vec{r}^{\prime}.
\end{equation}
Here $\vec{C}$ and $\phi$ at point $\vec{r}$ and time $t$ is calculated from magnetic current and charge distribution at distant position $\vec{r}^{\prime } $ at an earlier time $t^{ \prime }=t-\left|\vec{r}-\vec{r}^{\prime}\right|/c$ (known as the retarded time). The location $\vec{r}^{ \prime } $ is a source point within volume $\Omega$ that contains the magnetic current distribution. The integration variable, ${d} ^{3}\vec{r}^ { \prime }$, is a volume element around position $r^{\prime}$.  In a similar way the magnetic potentials may be written in terms of the electric current density and magnetic charge density, however, as such work focuses on voltage sources, we do not give the formulae but refer the reader to Refs \cite{RHbook2012,Balanis2012}.  

The two potential formulation also means we can separate  Maxwell's equations into two parts given by,
\begin{align}
&\epsilon_0\vec{\nabla}\cdot\vec{E}_A=\rho^i_e+\rho_f,\label{E1}\\
&\vec{\nabla} \times \vec{B}_A-\epsilon_0\mu_0\frac{\partial \vec{E}_A}{\partial t}=\mu_0(\vec{J}_e^i+\vec{J}_f),\label{E2}\\
&\vec{\nabla} \cdot \vec{B}_A=0,\label{E3}\\
&\vec{\nabla} \times \vec{E}_A+\frac{\partial \vec{B}_A}{\partial t}=0,\label{E4}
\end{align}
for the electric sources, and
\begin{align}
&\vec{\nabla}\cdot\vec{E}_C=0,\label{M1}\\
&\vec{\nabla} \times \vec{B}_C-\epsilon_0\mu_0\frac{\partial \vec{E}_C}{\partial t}=0,\label{M2}\\
&\vec{\nabla} \cdot \vec{B_C}=\rho_m^i,\label{M3}\\
&\vec{\nabla} \times \vec{E}_C+\frac{\partial \vec{B}_C}{\partial t}=-\vec{J}_m^i, \label{M4}
\end{align}
for the magnetic sources. Here, the impressed sources, $\rho^i_e$, $\vec{J}_e^i$, $\rho^i_m$ and $\vec{J}_m^i$, in equations (\ref{E1})$\rightarrow$(\ref{M4}) can only exist due to an impressed external source exciting the system. Also, in general, there may be some free charge and current in the system, $\rho_f$ and $\vec{J}_f$ respectively. Because magnetic monopoles do not exist, the effective magnetic current, $\vec{J}_{m}^i$, and the effective magnetic charge terms, $\rho^i_{m}$ can only exist as impressed sources. Equations (\ref{M1}) to (\ref{M4}) are the dual representation of equations (\ref{E1}) to (\ref{E4}). Due to the conservation of charge, impressed source currents and charges must satisfy the continuity equations,
\begin{equation}
\frac{\partial \rho^i_{e}}{\partial t}=-\vec{\nabla}\cdot\vec{J}^i_{e}, ~~\text{and}~~\frac{\partial \rho_{f}}{\partial t}=-\vec{\nabla}\cdot\vec{J}_{f}\label{J1}
\end{equation}
\begin{equation}
\frac{\partial \rho^i_{m}}{\partial t}=-\vec{\nabla}\cdot\vec{J}_{m}^i, \label{J2}
\end{equation}
which completes Maxwell's equations with impressed sources, which describe how electromagnetic energy or electricity can be generated from an external impressed energy source.


\begin{thebibliography}{59}%
\makeatletter
\providecommand \@ifxundefined [1]{%
 \@ifx{#1\undefined}
}%
\providecommand \@ifnum [1]{%
 \ifnum #1\expandafter \@firstoftwo
 \else \expandafter \@secondoftwo
 \fi
}%
\providecommand \@ifx [1]{%
 \ifx #1\expandafter \@firstoftwo
 \else \expandafter \@secondoftwo
 \fi
}%
\providecommand \natexlab [1]{#1}%
\providecommand \enquote  [1]{``#1''}%
\providecommand \bibnamefont  [1]{#1}%
\providecommand \bibfnamefont [1]{#1}%
\providecommand \citenamefont [1]{#1}%
\providecommand \href@noop [0]{\@secondoftwo}%
\providecommand \href [0]{\begingroup \@sanitize@url \@href}%
\providecommand \@href[1]{\@@startlink{#1}\@@href}%
\providecommand \@@href[1]{\endgroup#1\@@endlink}%
\providecommand \@sanitize@url [0]{\catcode `\\12\catcode `\$12\catcode
  `\&12\catcode `\#12\catcode `\^12\catcode `\_12\catcode `\%12\relax}%
\providecommand \@@startlink[1]{}%
\providecommand \@@endlink[0]{}%
\providecommand \url  [0]{\begingroup\@sanitize@url \@url }%
\providecommand \@url [1]{\endgroup\@href {#1}{\urlprefix }}%
\providecommand \urlprefix  [0]{URL }%
\providecommand \Eprint [0]{\href }%
\providecommand \doibase [0]{http://dx.doi.org/}%
\providecommand \selectlanguage [0]{\@gobble}%
\providecommand \bibinfo  [0]{\@secondoftwo}%
\providecommand \bibfield  [0]{\@secondoftwo}%
\providecommand \translation [1]{[#1]}%
\providecommand \BibitemOpen [0]{}%
\providecommand \bibitemStop [0]{}%
\providecommand \bibitemNoStop [0]{.\EOS\space}%
\providecommand \EOS [0]{\spacefactor3000\relax}%
\providecommand \BibitemShut  [1]{\csname bibitem#1\endcsname}%
\let\auto@bib@innerbib\@empty
\bibitem [{\citenamefont {Coleman}(1953)}]{Coleman53}%
  \BibitemOpen
  \bibfield  {author} {\bibinfo {author} {\bibfnamefont {J.~H.}\ \bibnamefont
  {Coleman}},\ }\bibfield  {title} {\enquote {\bibinfo {title} {Radioisotopic
  high-potential, low-current sources},}\ }\href@noop {} {\bibfield  {journal}
  {\bibinfo  {journal} {Nucleonics}\ }\textbf {\bibinfo {volume} {11}},\
  \bibinfo {pages} {42--25} (\bibinfo {year} {1953})}\BibitemShut {NoStop}%
\bibitem [{\citenamefont {Lal}\ and\ \citenamefont
  {Blanchard}(2004)}]{NucBatt}%
  \BibitemOpen
  \bibfield  {author} {\bibinfo {author} {\bibfnamefont {Amit}\ \bibnamefont
  {Lal}}\ and\ \bibinfo {author} {\bibfnamefont {James}\ \bibnamefont
  {Blanchard}},\ }\bibfield  {title} {\enquote {\bibinfo {title} {Nuclear
  batteries the daintiest dynamos},}\ }\href@noop {} {\bibfield  {journal}
  {\bibinfo  {journal} {IEEE Spectrum}\ }\textbf {\bibinfo {volume}
  {September}},\ \bibinfo {pages} {36 -- 41} (\bibinfo {year}
  {2004})}\BibitemShut {NoStop}%
\bibitem [{\citenamefont {Li}\ \emph {et~al.}(2002)\citenamefont {Li},
  \citenamefont {Lal}, \citenamefont {Blanchard},\ and\ \citenamefont
  {Henderson}}]{nucbat2002}%
  \BibitemOpen
  \bibfield  {author} {\bibinfo {author} {\bibfnamefont {Hui}\ \bibnamefont
  {Li}}, \bibinfo {author} {\bibfnamefont {Amit}\ \bibnamefont {Lal}}, \bibinfo
  {author} {\bibfnamefont {James}\ \bibnamefont {Blanchard}}, \ and\ \bibinfo
  {author} {\bibfnamefont {Douglass}\ \bibnamefont {Henderson}},\ }\bibfield
  {title} {\enquote {\bibinfo {title} {Self-reciprocating radioisotope-powered
  cantilever},}\ }\href {\doibase 10.1063/1.1479755} {\bibfield  {journal}
  {\bibinfo  {journal} {Journal of Applied Physics}\ }\textbf {\bibinfo
  {volume} {92}},\ \bibinfo {pages} {1122--1127} (\bibinfo {year} {2002})},\
  \Eprint {http://arxiv.org/abs/https://doi.org/10.1063/1.1479755}
  {https://doi.org/10.1063/1.1479755} \BibitemShut {NoStop}%
\bibitem [{\citenamefont {Li}\ \emph {et~al.}(2001)\citenamefont {Li},
  \citenamefont {Lal}, \citenamefont {Blanchard},\ and\ \citenamefont
  {Henderson}}]{Li2001}%
  \BibitemOpen
  \bibfield  {author} {\bibinfo {author} {\bibfnamefont {H.}~\bibnamefont
  {Li}}, \bibinfo {author} {\bibfnamefont {A}~\bibnamefont {Lal}}, \bibinfo
  {author} {\bibfnamefont {J.}~\bibnamefont {Blanchard}}, \ and\ \bibinfo
  {author} {\bibfnamefont {D.}~\bibnamefont {Henderson}},\ }\bibfield  {title}
  {\enquote {\bibinfo {title} {Self-reciprocating radioisotope-powered
  cantilever},}\ }in\ \href@noop {} {\emph {\bibinfo {booktitle} {Transducers
  '01 Eurosensors XV}}},\ \bibinfo {editor} {edited by\ \bibinfo {editor}
  {\bibfnamefont {E.}~\bibnamefont {Obermeier}}}\ (\bibinfo {year}
  {2001})\BibitemShut {NoStop}%
\bibitem [{\citenamefont {Wang}\ \emph {et~al.}(2017)\citenamefont {Wang},
  \citenamefont {Jiang},\ and\ \citenamefont {Xu}}]{WANG20179}%
  \BibitemOpen
  \bibfield  {author} {\bibinfo {author} {\bibfnamefont {Zhong~Lin}\
  \bibnamefont {Wang}}, \bibinfo {author} {\bibfnamefont {Tao}\ \bibnamefont
  {Jiang}}, \ and\ \bibinfo {author} {\bibfnamefont {Liang}\ \bibnamefont
  {Xu}},\ }\bibfield  {title} {\enquote {\bibinfo {title} {Toward the blue
  energy dream by triboelectric nanogenerator networks},}\ }\href {\doibase
  https://doi.org/10.1016/j.nanoen.2017.06.035} {\bibfield  {journal} {\bibinfo
   {journal} {Nano Energy}\ }\textbf {\bibinfo {volume} {39}},\ \bibinfo
  {pages} {9 -- 23} (\bibinfo {year} {2017})}\BibitemShut {NoStop}%
\bibitem [{\citenamefont {Wang}(2017)}]{WANG201774}%
  \BibitemOpen
  \bibfield  {author} {\bibinfo {author} {\bibfnamefont {Zhong~Lin}\
  \bibnamefont {Wang}},\ }\bibfield  {title} {\enquote {\bibinfo {title} {On
  maxwell's displacement current for energy and sensors: the origin of
  nanogenerators},}\ }\href {\doibase
  https://doi.org/10.1016/j.mattod.2016.12.001} {\bibfield  {journal} {\bibinfo
   {journal} {Materials Today}\ }\textbf {\bibinfo {volume} {20}},\ \bibinfo
  {pages} {74 -- 82} (\bibinfo {year} {2017})}\BibitemShut {NoStop}%
\bibitem [{\citenamefont {Sessler}\ \emph {et~al.}(2016)\citenamefont
  {Sessler}, \citenamefont {Pondrom},\ and\ \citenamefont
  {Zhang}}]{Sessler2016}%
  \BibitemOpen
  \bibfield  {author} {\bibinfo {author} {\bibfnamefont {G.~M.}\ \bibnamefont
  {Sessler}}, \bibinfo {author} {\bibfnamefont {P.}~\bibnamefont {Pondrom}}, \
  and\ \bibinfo {author} {\bibfnamefont {X.}~\bibnamefont {Zhang}},\ }\bibfield
   {title} {\enquote {\bibinfo {title} {Stacked and folded piezoelectrets for
  vibration-based energy harvesting},}\ }\href {\doibase
  10.1080/01411594.2016.1202408} {\bibfield  {journal} {\bibinfo  {journal}
  {Phase Transitions}\ }\textbf {\bibinfo {volume} {89}},\ \bibinfo {pages}
  {667--677} (\bibinfo {year} {2016})},\ \Eprint
  {http://arxiv.org/abs/https://doi.org/10.1080/01411594.2016.1202408}
  {https://doi.org/10.1080/01411594.2016.1202408} \BibitemShut {NoStop}%
\bibitem [{\citenamefont {Wang}\ and\ \citenamefont {Song}(2006)}]{Wang242}%
  \BibitemOpen
  \bibfield  {author} {\bibinfo {author} {\bibfnamefont {Zhong~Lin}\
  \bibnamefont {Wang}}\ and\ \bibinfo {author} {\bibfnamefont {Jinhui}\
  \bibnamefont {Song}},\ }\bibfield  {title} {\enquote {\bibinfo {title}
  {Piezoelectric nanogenerators based on zinc oxide nanowire arrays},}\ }\href
  {\doibase 10.1126/science.1124005} {\bibfield  {journal} {\bibinfo  {journal}
  {Science}\ }\textbf {\bibinfo {volume} {312}},\ \bibinfo {pages} {242--246}
  (\bibinfo {year} {2006})}\BibitemShut {NoStop}%
\bibitem [{\citenamefont {Yang}\ \emph {et~al.}(2009)\citenamefont {Yang},
  \citenamefont {Qin}, \citenamefont {Dai},\ and\ \citenamefont
  {Wang}}]{Yang:2009wa}%
  \BibitemOpen
  \bibfield  {author} {\bibinfo {author} {\bibfnamefont {Rusen}\ \bibnamefont
  {Yang}}, \bibinfo {author} {\bibfnamefont {Yong}\ \bibnamefont {Qin}},
  \bibinfo {author} {\bibfnamefont {Liming}\ \bibnamefont {Dai}}, \ and\
  \bibinfo {author} {\bibfnamefont {Zhong~Lin}\ \bibnamefont {Wang}},\
  }\bibfield  {title} {\enquote {\bibinfo {title} {Power generation with
  laterally packaged piezoelectric fine wires},}\ }\href {\doibase
  10.1038/nnano.2008.314} {\bibfield  {journal} {\bibinfo  {journal} {Nature
  Nanotechnology}\ }\textbf {\bibinfo {volume} {4}},\ \bibinfo {pages} {34--39}
  (\bibinfo {year} {2009})}\BibitemShut {NoStop}%
\bibitem [{\citenamefont {Fan}\ \emph {et~al.}(2012)\citenamefont {Fan},
  \citenamefont {Tian},\ and\ \citenamefont {{Lin Wang}}}]{FAN2012328}%
  \BibitemOpen
  \bibfield  {author} {\bibinfo {author} {\bibfnamefont {Feng-Ru}\ \bibnamefont
  {Fan}}, \bibinfo {author} {\bibfnamefont {Zhong-Qun}\ \bibnamefont {Tian}}, \
  and\ \bibinfo {author} {\bibfnamefont {Zhong}\ \bibnamefont {{Lin Wang}}},\
  }\bibfield  {title} {\enquote {\bibinfo {title} {Flexible triboelectric
  generator},}\ }\href {\doibase https://doi.org/10.1016/j.nanoen.2012.01.004}
  {\bibfield  {journal} {\bibinfo  {journal} {Nano Energy}\ }\textbf {\bibinfo
  {volume} {1}},\ \bibinfo {pages} {328 -- 334} (\bibinfo {year}
  {2012})}\BibitemShut {NoStop}%
\bibitem [{\citenamefont {Wang}(2013)}]{Wang2013}%
  \BibitemOpen
  \bibfield  {author} {\bibinfo {author} {\bibfnamefont {Zhong~Lin}\
  \bibnamefont {Wang}},\ }\bibfield  {title} {\enquote {\bibinfo {title}
  {Triboelectric nanogenerators as new energy technology for self-powered
  systems and as active mechanical and chemical sensors},}\ }\href {\doibase
  10.1021/nn404614z} {\bibfield  {journal} {\bibinfo  {journal} {ACS Nano}\
  }\textbf {\bibinfo {volume} {7}},\ \bibinfo {pages} {9533--9557} (\bibinfo
  {year} {2013})},\ \bibinfo {note} {pMID: 24079963},\ \Eprint
  {http://arxiv.org/abs/https://doi.org/10.1021/nn404614z}
  {https://doi.org/10.1021/nn404614z} \BibitemShut {NoStop}%
\bibitem [{\citenamefont {Xue}\ \emph {et~al.}(2017)\citenamefont {Xue},
  \citenamefont {Yang}, \citenamefont {Wang}, \citenamefont {Luo},
  \citenamefont {Wang}, \citenamefont {Lin}, \citenamefont {Liang},\ and\
  \citenamefont {Luo}}]{XUE2017147}%
  \BibitemOpen
  \bibfield  {author} {\bibinfo {author} {\bibfnamefont {Hao}\ \bibnamefont
  {Xue}}, \bibinfo {author} {\bibfnamefont {Quan}\ \bibnamefont {Yang}},
  \bibinfo {author} {\bibfnamefont {Dingyi}\ \bibnamefont {Wang}}, \bibinfo
  {author} {\bibfnamefont {Weijian}\ \bibnamefont {Luo}}, \bibinfo {author}
  {\bibfnamefont {Wenqian}\ \bibnamefont {Wang}}, \bibinfo {author}
  {\bibfnamefont {Mushun}\ \bibnamefont {Lin}}, \bibinfo {author}
  {\bibfnamefont {Dingli}\ \bibnamefont {Liang}}, \ and\ \bibinfo {author}
  {\bibfnamefont {Qiming}\ \bibnamefont {Luo}},\ }\bibfield  {title} {\enquote
  {\bibinfo {title} {A wearable pyroelectric nanogenerator and self-powered
  breathing sensor},}\ }\href {\doibase
  https://doi.org/10.1016/j.nanoen.2017.05.056} {\bibfield  {journal} {\bibinfo
   {journal} {Nano Energy}\ }\textbf {\bibinfo {volume} {38}},\ \bibinfo
  {pages} {147 -- 154} (\bibinfo {year} {2017})}\BibitemShut {NoStop}%
\bibitem [{\citenamefont {Yang}\ \emph
  {et~al.}(2012{\natexlab{a}})\citenamefont {Yang}, \citenamefont {Guo},
  \citenamefont {Pradel}, \citenamefont {Zhu}, \citenamefont {Zhou},
  \citenamefont {Zhang}, \citenamefont {Hu}, \citenamefont {Lin},\ and\
  \citenamefont {Wang}}]{Yang20122833}%
  \BibitemOpen
  \bibfield  {author} {\bibinfo {author} {\bibfnamefont {Y.}~\bibnamefont
  {Yang}}, \bibinfo {author} {\bibfnamefont {W.}~\bibnamefont {Guo}}, \bibinfo
  {author} {\bibfnamefont {K.C.}\ \bibnamefont {Pradel}}, \bibinfo {author}
  {\bibfnamefont {G.}~\bibnamefont {Zhu}}, \bibinfo {author} {\bibfnamefont
  {Y.}~\bibnamefont {Zhou}}, \bibinfo {author} {\bibfnamefont {Y.}~\bibnamefont
  {Zhang}}, \bibinfo {author} {\bibfnamefont {Y.}~\bibnamefont {Hu}}, \bibinfo
  {author} {\bibfnamefont {L.}~\bibnamefont {Lin}}, \ and\ \bibinfo {author}
  {\bibfnamefont {Z.L.}\ \bibnamefont {Wang}},\ }\bibfield  {title} {\enquote
  {\bibinfo {title} {Pyroelectric nanogenerators for harvesting thermoelectric
  energy},}\ }\href {\doibase 10.1021/nl3003039} {\bibfield  {journal}
  {\bibinfo  {journal} {Nano Letters}\ }\textbf {\bibinfo {volume} {12}},\
  \bibinfo {pages} {2833--2838} (\bibinfo {year} {2012}{\natexlab{a}})},\
  \bibinfo {note} {cited By 384}\BibitemShut {NoStop}%
\bibitem [{\citenamefont {Ko}\ \emph {et~al.}(2016)\citenamefont {Ko},
  \citenamefont {Kim}, \citenamefont {Won}, \citenamefont {Ahn}, \citenamefont
  {Kim}, \citenamefont {Kingon}, \citenamefont {Kim}, \citenamefont {Ko},\ and\
  \citenamefont {Jung}}]{Ko20166504}%
  \BibitemOpen
  \bibfield  {author} {\bibinfo {author} {\bibfnamefont {Y.J.}\ \bibnamefont
  {Ko}}, \bibinfo {author} {\bibfnamefont {D.Y.}\ \bibnamefont {Kim}}, \bibinfo
  {author} {\bibfnamefont {S.S.}\ \bibnamefont {Won}}, \bibinfo {author}
  {\bibfnamefont {C.W.}\ \bibnamefont {Ahn}}, \bibinfo {author} {\bibfnamefont
  {I.W.}\ \bibnamefont {Kim}}, \bibinfo {author} {\bibfnamefont {A.I.}\
  \bibnamefont {Kingon}}, \bibinfo {author} {\bibfnamefont {S.-H.}\
  \bibnamefont {Kim}}, \bibinfo {author} {\bibfnamefont {J.-H.}\ \bibnamefont
  {Ko}}, \ and\ \bibinfo {author} {\bibfnamefont {J.H.}\ \bibnamefont {Jung}},\
  }\bibfield  {title} {\enquote {\bibinfo {title} {Flexible pb(zr0.52ti0.48)o3
  films for a hybrid piezoelectric-pyroelectric nanogenerator under harsh
  environments},}\ }\href {\doibase 10.1021/acsami.6b00054} {\bibfield
  {journal} {\bibinfo  {journal} {ACS Applied Materials and Interfaces}\
  }\textbf {\bibinfo {volume} {8}},\ \bibinfo {pages} {6504--6511} (\bibinfo
  {year} {2016})},\ \bibinfo {note} {cited By 32}\BibitemShut {NoStop}%
\bibitem [{\citenamefont {Yang}\ \emph
  {et~al.}(2012{\natexlab{b}})\citenamefont {Yang}, \citenamefont {Jung},
  \citenamefont {Yun}, \citenamefont {Zhang}, \citenamefont {Pradel},
  \citenamefont {Guo},\ and\ \citenamefont {Wang}}]{Yang20125357}%
  \BibitemOpen
  \bibfield  {author} {\bibinfo {author} {\bibfnamefont {Y.}~\bibnamefont
  {Yang}}, \bibinfo {author} {\bibfnamefont {J.H.}\ \bibnamefont {Jung}},
  \bibinfo {author} {\bibfnamefont {B.K.}\ \bibnamefont {Yun}}, \bibinfo
  {author} {\bibfnamefont {F.}~\bibnamefont {Zhang}}, \bibinfo {author}
  {\bibfnamefont {K.C.}\ \bibnamefont {Pradel}}, \bibinfo {author}
  {\bibfnamefont {W.}~\bibnamefont {Guo}}, \ and\ \bibinfo {author}
  {\bibfnamefont {Z.L.}\ \bibnamefont {Wang}},\ }\bibfield  {title} {\enquote
  {\bibinfo {title} {Flexible pyroelectric nanogenerators using a composite
  structure of lead-free knbo3 nanowires},}\ }\href {\doibase
  10.1002/adma.201201414} {\bibfield  {journal} {\bibinfo  {journal} {Advanced
  Materials}\ }\textbf {\bibinfo {volume} {24}},\ \bibinfo {pages} {5357--5362}
  (\bibinfo {year} {2012}{\natexlab{b}})},\ \bibinfo {note} {cited By
  145}\BibitemShut {NoStop}%
\bibitem [{\citenamefont {Zi}\ \emph {et~al.}(2015)\citenamefont {Zi},
  \citenamefont {Lin}, \citenamefont {Wang}, \citenamefont {Wang},
  \citenamefont {Chen}, \citenamefont {Fan}, \citenamefont {Yang},
  \citenamefont {Yi},\ and\ \citenamefont {Wang}}]{Zi20152340}%
  \BibitemOpen
  \bibfield  {author} {\bibinfo {author} {\bibfnamefont {Y.}~\bibnamefont
  {Zi}}, \bibinfo {author} {\bibfnamefont {L.}~\bibnamefont {Lin}}, \bibinfo
  {author} {\bibfnamefont {J.}~\bibnamefont {Wang}}, \bibinfo {author}
  {\bibfnamefont {S.}~\bibnamefont {Wang}}, \bibinfo {author} {\bibfnamefont
  {J.}~\bibnamefont {Chen}}, \bibinfo {author} {\bibfnamefont {X.}~\bibnamefont
  {Fan}}, \bibinfo {author} {\bibfnamefont {P.-K.}\ \bibnamefont {Yang}},
  \bibinfo {author} {\bibfnamefont {F.}~\bibnamefont {Yi}}, \ and\ \bibinfo
  {author} {\bibfnamefont {Z.L.}\ \bibnamefont {Wang}},\ }\bibfield  {title}
  {\enquote {\bibinfo {title} {Triboelectric-pyroelectric-piezoelectric hybrid
  cell for high-efficiency energy-harvesting and self-powered sensing},}\
  }\href {\doibase 10.1002/adma.201500121} {\bibfield  {journal} {\bibinfo
  {journal} {Advanced Materials}\ }\textbf {\bibinfo {volume} {27}},\ \bibinfo
  {pages} {2340--2347} (\bibinfo {year} {2015})}\BibitemShut {NoStop}%
\bibitem [{\citenamefont {Lee}\ \emph {et~al.}(2014)\citenamefont {Lee},
  \citenamefont {Lee}, \citenamefont {Gupta}, \citenamefont {Kim},
  \citenamefont {Lee}, \citenamefont {Oh}, \citenamefont {Ryu}, \citenamefont
  {Yoo}, \citenamefont {Kang}, \citenamefont {Yoon}, \citenamefont {Yoo},\ and\
  \citenamefont {Kim}}]{Lee2014765}%
  \BibitemOpen
  \bibfield  {author} {\bibinfo {author} {\bibfnamefont {J.-H.}\ \bibnamefont
  {Lee}}, \bibinfo {author} {\bibfnamefont {K.Y.}\ \bibnamefont {Lee}},
  \bibinfo {author} {\bibfnamefont {M.K.}\ \bibnamefont {Gupta}}, \bibinfo
  {author} {\bibfnamefont {T.Y.}\ \bibnamefont {Kim}}, \bibinfo {author}
  {\bibfnamefont {D.-Y.}\ \bibnamefont {Lee}}, \bibinfo {author} {\bibfnamefont
  {J.}~\bibnamefont {Oh}}, \bibinfo {author} {\bibfnamefont {C.}~\bibnamefont
  {Ryu}}, \bibinfo {author} {\bibfnamefont {W.J.}\ \bibnamefont {Yoo}},
  \bibinfo {author} {\bibfnamefont {C.-Y.}\ \bibnamefont {Kang}}, \bibinfo
  {author} {\bibfnamefont {S.-J.}\ \bibnamefont {Yoon}}, \bibinfo {author}
  {\bibfnamefont {J.-B.}\ \bibnamefont {Yoo}}, \ and\ \bibinfo {author}
  {\bibfnamefont {S.-W.}\ \bibnamefont {Kim}},\ }\bibfield  {title} {\enquote
  {\bibinfo {title} {Highly stretchable piezoelectric-pyroelectric hybrid
  nanogenerator},}\ }\href {\doibase 10.1002/adma.201303570} {\bibfield
  {journal} {\bibinfo  {journal} {Advanced Materials}\ }\textbf {\bibinfo
  {volume} {26}},\ \bibinfo {pages} {765--769} (\bibinfo {year} {2014})},\
  \bibinfo {note} {cited By 290}\BibitemShut {NoStop}%
\bibitem [{\citenamefont {Park}\ \emph {et~al.}(2015)\citenamefont {Park},
  \citenamefont {Na}, \citenamefont {Kim}, \citenamefont {Kim}, \citenamefont
  {Shin},\ and\ \citenamefont {Kim}}]{Park201511830}%
  \BibitemOpen
  \bibfield  {author} {\bibinfo {author} {\bibfnamefont {T.}~\bibnamefont
  {Park}}, \bibinfo {author} {\bibfnamefont {J.}~\bibnamefont {Na}}, \bibinfo
  {author} {\bibfnamefont {B.}~\bibnamefont {Kim}}, \bibinfo {author}
  {\bibfnamefont {Y.}~\bibnamefont {Kim}}, \bibinfo {author} {\bibfnamefont
  {H.}~\bibnamefont {Shin}}, \ and\ \bibinfo {author} {\bibfnamefont
  {E.}~\bibnamefont {Kim}},\ }\bibfield  {title} {\enquote {\bibinfo {title}
  {Photothermally activated pyroelectric polymer films for harvesting of solar
  heat with a hybrid energy cell structure},}\ }\href {\doibase
  10.1021/acsnano.5b04042} {\bibfield  {journal} {\bibinfo  {journal} {ACS
  Nano}\ }\textbf {\bibinfo {volume} {9}},\ \bibinfo {pages} {11830--11839}
  (\bibinfo {year} {2015})},\ \bibinfo {note} {cited By 46}\BibitemShut
  {NoStop}%
\bibitem [{\citenamefont {{Popovic}}(1981)}]{Popovic81}%
  \BibitemOpen
  \bibfield  {author} {\bibinfo {author} {\bibfnamefont {B.~D.}\ \bibnamefont
  {{Popovic}}},\ }\bibfield  {title} {\enquote {\bibinfo {title}
  {Electromagnetic field theorems},}\ }\href {\doibase
  10.1049/ip-a-1.1981.0007} {\bibfield  {journal} {\bibinfo  {journal} {IEE
  Proceedings A - Physical Science, Measurement and Instrumentation, Management
  and Education - Reviews}\ }\textbf {\bibinfo {volume} {128}},\ \bibinfo
  {pages} {47--63} (\bibinfo {year} {1981})}\BibitemShut {NoStop}%
\bibitem [{\citenamefont {Harrington}(2012)}]{RHbook2012}%
  \BibitemOpen
  \bibfield  {author} {\bibinfo {author} {\bibfnamefont {Roger~E.}\
  \bibnamefont {Harrington}},\ }\href@noop {} {\emph {\bibinfo {title}
  {Introduction to Electromagnetic Engineering}}},\ \bibinfo {edition} {2nd}\
  ed.\ (\bibinfo  {publisher} {Dover Publications, Inc.},\ \bibinfo {address}
  {31 East 2nd Street, Mineola, NY 11501},\ \bibinfo {year} {2012})\BibitemShut
  {NoStop}%
\bibitem [{\citenamefont {Balanis}(2012)}]{Balanis2012}%
  \BibitemOpen
  \bibfield  {author} {\bibinfo {author} {\bibfnamefont {Constantine~A}\
  \bibnamefont {Balanis}},\ }\href@noop {} {\emph {\bibinfo {title} {Advanced
  Engineering Electromagnetics}}}\ (\bibinfo  {publisher} {John Wiley,},\
  \bibinfo {year} {2012})\BibitemShut {NoStop}%
\bibitem [{\citenamefont {Jordan}\ and\ \citenamefont {Balmain}(1968)}]{ECJ}%
  \BibitemOpen
  \bibfield  {author} {\bibinfo {author} {\bibfnamefont {Edward~Conrad}\
  \bibnamefont {Jordan}}\ and\ \bibinfo {author} {\bibfnamefont {Kieth~G.}\
  \bibnamefont {Balmain}},\ }\enquote {\bibinfo {title} {Electromagnetic waves
  and radiating systems},}\ \ (\bibinfo  {publisher} {Prentice Hall, Inc.},\
  \bibinfo {year} {1968})\ Chap.~\bibinfo {chapter} {13},\ \bibinfo {edition}
  {2nd}\ ed.\BibitemShut {Stop}%
\bibitem [{\citenamefont {D.Monteath}(1951)}]{Monteath51}%
  \BibitemOpen
  \bibfield  {author} {\bibinfo {author} {\bibfnamefont {G.}~\bibnamefont
  {D.Monteath}},\ }\bibfield  {title} {\enquote {\bibinfo {title} {Application
  of the compensation theorem to certain radiation and propagation problems},}\
  }\href@noop {} {\bibfield  {journal} {\bibinfo  {journal} {Proc. IEE}\
  }\textbf {\bibinfo {volume} {98 Pt. IV}},\ \bibinfo {pages} {23--30}
  (\bibinfo {year} {1951})}\BibitemShut {NoStop}%
\bibitem [{\citenamefont {Stumpf}(2018)}]{ERAntTh18}%
  \BibitemOpen
  \bibfield  {author} {\bibinfo {author} {\bibfnamefont {Martin}\ \bibnamefont
  {Stumpf}},\ }\href@noop {} {\emph {\bibinfo {title} {Electromagnetic
  Reciprocity in Antenna Theory}}},\ edited by\ \bibinfo {editor}
  {\bibfnamefont {Tariq}\ \bibnamefont {Samad}}\ (\bibinfo  {publisher} {IEEE
  Press},\ \bibinfo {year} {2018})\BibitemShut {NoStop}%
\bibitem [{\citenamefont {Tobar}\ \emph {et~al.}(2019)\citenamefont {Tobar},
  \citenamefont {McAllister},\ and\ \citenamefont {Goryachev}}]{TobarModAx19}%
  \BibitemOpen
  \bibfield  {author} {\bibinfo {author} {\bibfnamefont {Michael~E.}\
  \bibnamefont {Tobar}}, \bibinfo {author} {\bibfnamefont {Ben~T.}\
  \bibnamefont {McAllister}}, \ and\ \bibinfo {author} {\bibfnamefont {Maxim}\
  \bibnamefont {Goryachev}},\ }\bibfield  {title} {\enquote {\bibinfo {title}
  {Modified axion electrodynamics as impressed electromagnetic sources through
  oscillating background polarization and magnetization},}\ }\href {\doibase
  https://doi.org/10.1016/j.dark.2019.100339} {\bibfield  {journal} {\bibinfo
  {journal} {Physics of the Dark Universe}\ }\textbf {\bibinfo {volume} {26}},\
  \bibinfo {pages} {100339} (\bibinfo {year} {2019})}\BibitemShut {NoStop}%
\bibitem [{\citenamefont {Tobar}\ \emph {et~al.}(2020)\citenamefont {Tobar},
  \citenamefont {McAllister},\ and\ \citenamefont
  {Goryachev}}]{TOBAR2020100624}%
  \BibitemOpen
  \bibfield  {author} {\bibinfo {author} {\bibfnamefont {Michael~E.}\
  \bibnamefont {Tobar}}, \bibinfo {author} {\bibfnamefont {Ben~T.}\
  \bibnamefont {McAllister}}, \ and\ \bibinfo {author} {\bibfnamefont {Maxim}\
  \bibnamefont {Goryachev}},\ }\bibfield  {title} {\enquote {\bibinfo {title}
  {Broadband electrical action sensing techniques with conducting wires for
  low-mass dark matter axion detection},}\ }\href {\doibase
  https://doi.org/10.1016/j.dark.2020.100624} {\bibfield  {journal} {\bibinfo
  {journal} {Physics of the Dark Universe}\ }\textbf {\bibinfo {volume} {30}},\
  \bibinfo {pages} {100624} (\bibinfo {year} {2020})}\BibitemShut {NoStop}%
\bibitem [{\citenamefont {Cao}\ and\ \citenamefont
  {Zhitnitsky}(2017)}]{Cao2017}%
  \BibitemOpen
  \bibfield  {author} {\bibinfo {author} {\bibfnamefont {ChunJun}\ \bibnamefont
  {Cao}}\ and\ \bibinfo {author} {\bibfnamefont {Ariel}\ \bibnamefont
  {Zhitnitsky}},\ }\bibfield  {title} {\enquote {\bibinfo {title} {Axion
  detection via topological casimir effect},}\ }\href {\doibase
  10.1103/PhysRevD.96.015013} {\bibfield  {journal} {\bibinfo  {journal} {Phys.
  Rev. D}\ }\textbf {\bibinfo {volume} {96}},\ \bibinfo {pages} {015013}
  (\bibinfo {year} {2017})}\BibitemShut {NoStop}%
\bibitem [{\citenamefont {Griffiths}(1999)}]{GriffBook}%
  \BibitemOpen
  \bibfield  {author} {\bibinfo {author} {\bibfnamefont {David~J}\ \bibnamefont
  {Griffiths}},\ }\href@noop {} {\emph {\bibinfo {title} {Introduction to
  Electrodynamics}}},\ \bibinfo {edition} {3rd}\ ed.\ (\bibinfo  {publisher}
  {Prentice Hall},\ \bibinfo {address} {Upper Saddle River, New Jersey 07458},\
  \bibinfo {year} {1999})\BibitemShut {NoStop}%
\bibitem [{\citenamefont {Roberts}(1983)}]{Roberts1983}%
  \BibitemOpen
  \bibfield  {author} {\bibinfo {author} {\bibfnamefont {Dana}\ \bibnamefont
  {Roberts}},\ }\bibfield  {title} {\enquote {\bibinfo {title} {How batteries
  work: A gravitational analog},}\ }\href {\doibase 10.1119/1.13128} {\bibfield
   {journal} {\bibinfo  {journal} {American Journal of Physics}\ }\textbf
  {\bibinfo {volume} {51}},\ \bibinfo {pages} {829--831} (\bibinfo {year}
  {1983})},\ \Eprint {http://arxiv.org/abs/https://doi.org/10.1119/1.13128}
  {https://doi.org/10.1119/1.13128} \BibitemShut {NoStop}%
\bibitem [{\citenamefont {Baierlein}(2001)}]{Baierlein2001}%
  \BibitemOpen
  \bibfield  {author} {\bibinfo {author} {\bibfnamefont {Ralph}\ \bibnamefont
  {Baierlein}},\ }\bibfield  {title} {\enquote {\bibinfo {title} {The elusive
  chemical potential},}\ }\href {\doibase 10.1119/1.1336839} {\bibfield
  {journal} {\bibinfo  {journal} {American Journal of Physics}\ }\textbf
  {\bibinfo {volume} {69}},\ \bibinfo {pages} {423--434} (\bibinfo {year}
  {2001})},\ \Eprint {http://arxiv.org/abs/https://doi.org/10.1119/1.1336839}
  {https://doi.org/10.1119/1.1336839} \BibitemShut {NoStop}%
\bibitem [{\citenamefont {Saslow}(1999)}]{Saslow1999}%
  \BibitemOpen
  \bibfield  {author} {\bibinfo {author} {\bibfnamefont {Wayne~M.}\
  \bibnamefont {Saslow}},\ }\bibfield  {title} {\enquote {\bibinfo {title}
  {Voltaic cells for physicists: Two surface pumps and an internal
  resistance},}\ }\href {\doibase 10.1119/1.19327} {\bibfield  {journal}
  {\bibinfo  {journal} {American Journal of Physics}\ }\textbf {\bibinfo
  {volume} {67}},\ \bibinfo {pages} {574--583} (\bibinfo {year} {1999})},\
  \Eprint {http://arxiv.org/abs/https://doi.org/10.1119/1.19327}
  {https://doi.org/10.1119/1.19327} \BibitemShut {NoStop}%
\bibitem [{\citenamefont {Hum}(2020)}]{Hum20}%
  \BibitemOpen
  \bibfield  {author} {\bibinfo {author} {\bibfnamefont {Sean~Victor}\
  \bibnamefont {Hum}},\ }\href@noop {} {\emph {\bibinfo {title} {Ideal
  (Hertzian) Dipole}}},\ Vol.\ \bibinfo {volume} {ece422}\ (\bibinfo
  {publisher} {University of Toronto},\ \bibinfo {year} {2020})\BibitemShut
  {NoStop}%
\bibitem [{\citenamefont {Sessler}(1987)}]{Electrets}%
  \BibitemOpen
  \bibfield  {author} {\bibinfo {author} {\bibfnamefont {G.~M.}\ \bibnamefont
  {Sessler}},\ }\href@noop {} {\emph {\bibinfo {title} {Electrets}}}\ (\bibinfo
   {publisher} {Springer-Verlag, New York, Berlin, Heidelberg},\ \bibinfo
  {year} {1987})\BibitemShut {NoStop}%
\bibitem [{\citenamefont {Jefimenko}\ and\ \citenamefont
  {Walker}(1980)}]{Jefimenko1980}%
  \BibitemOpen
  \bibfield  {author} {\bibinfo {author} {\bibfnamefont {Oleg~D.}\ \bibnamefont
  {Jefimenko}}\ and\ \bibinfo {author} {\bibfnamefont {David~K.}\ \bibnamefont
  {Walker}},\ }\bibfield  {title} {\enquote {\bibinfo {title} {Electrets},}\
  }\href {\doibase 10.1119/1.2340651} {\bibfield  {journal} {\bibinfo
  {journal} {The Physics Teacher}\ }\textbf {\bibinfo {volume} {18}},\ \bibinfo
  {pages} {651--659} (\bibinfo {year} {1980})},\ \Eprint
  {http://arxiv.org/abs/https://doi.org/10.1119/1.2340651}
  {https://doi.org/10.1119/1.2340651} \BibitemShut {NoStop}%
\bibitem [{\citenamefont {Monkman}\ \emph {et~al.}(2017)\citenamefont
  {Monkman}, \citenamefont {Sindersberger}, \citenamefont {Diermeier},\ and\
  \citenamefont {Prem}}]{Monkman_2017}%
  \BibitemOpen
  \bibfield  {author} {\bibinfo {author} {\bibfnamefont {G~J}\ \bibnamefont
  {Monkman}}, \bibinfo {author} {\bibfnamefont {D}~\bibnamefont
  {Sindersberger}}, \bibinfo {author} {\bibfnamefont {A}~\bibnamefont
  {Diermeier}}, \ and\ \bibinfo {author} {\bibfnamefont {N}~\bibnamefont
  {Prem}},\ }\bibfield  {title} {\enquote {\bibinfo {title} {The magnetoactive
  electret},}\ }\href {\doibase 10.1088/1361-665x/aa738f} {\bibfield  {journal}
  {\bibinfo  {journal} {Smart Materials and Structures}\ }\textbf {\bibinfo
  {volume} {26}},\ \bibinfo {pages} {075010} (\bibinfo {year}
  {2017})}\BibitemShut {NoStop}%
\bibitem [{\citenamefont {Asanuma}\ \emph {et~al.}(2013)\citenamefont
  {Asanuma}, \citenamefont {Oguchi}, \citenamefont {Hara}, \citenamefont
  {Yoshida},\ and\ \citenamefont {Kuwano}}]{Asanuma2013}%
  \BibitemOpen
  \bibfield  {author} {\bibinfo {author} {\bibfnamefont {Haruhiko}\
  \bibnamefont {Asanuma}}, \bibinfo {author} {\bibfnamefont {Hiroyuki}\
  \bibnamefont {Oguchi}}, \bibinfo {author} {\bibfnamefont {Motoaki}\
  \bibnamefont {Hara}}, \bibinfo {author} {\bibfnamefont {Ryo}\ \bibnamefont
  {Yoshida}}, \ and\ \bibinfo {author} {\bibfnamefont {Hiroki}\ \bibnamefont
  {Kuwano}},\ }\bibfield  {title} {\enquote {\bibinfo {title} {Ferroelectric
  dipole electrets for output power enhancement in electrostatic vibration
  energy harvesters},}\ }\href {\doibase 10.1063/1.4824831} {\bibfield
  {journal} {\bibinfo  {journal} {Applied Physics Letters}\ }\textbf {\bibinfo
  {volume} {103}},\ \bibinfo {pages} {162901} (\bibinfo {year} {2013})},\
  \Eprint {http://arxiv.org/abs/https://doi.org/10.1063/1.4824831}
  {https://doi.org/10.1063/1.4824831} \BibitemShut {NoStop}%
\bibitem [{\citenamefont {Graz}\ and\ \citenamefont
  {Mellinger}(2016)}]{Graz2016}%
  \BibitemOpen
  \bibfield  {author} {\bibinfo {author} {\bibfnamefont {Ingrid}\ \bibnamefont
  {Graz}}\ and\ \bibinfo {author} {\bibfnamefont {Axel}\ \bibnamefont
  {Mellinger}},\ }\enquote {\bibinfo {title} {Polymer electrets and
  ferroelectrets as eaps: Fundamentals},}\ in\ \href {\doibase
  10.1007/978-3-319-31767-0_24-1} {\emph {\bibinfo {booktitle}
  {Electromechanically Active Polymers: A Concise Reference}}},\ \bibinfo
  {editor} {edited by\ \bibinfo {editor} {\bibfnamefont {Federico}\
  \bibnamefont {Carpi}}}\ (\bibinfo  {publisher} {Springer International
  Publishing},\ \bibinfo {address} {Cham},\ \bibinfo {year} {2016})\ pp.\
  \bibinfo {pages} {1--10}\BibitemShut {NoStop}%
\bibitem [{\citenamefont {Wan}\ and\ \citenamefont {Bowen}(2017)}]{C6TA09590A}%
  \BibitemOpen
  \bibfield  {author} {\bibinfo {author} {\bibfnamefont {Chaoying}\
  \bibnamefont {Wan}}\ and\ \bibinfo {author} {\bibfnamefont
  {Christopher~Rhys}\ \bibnamefont {Bowen}},\ }\bibfield  {title} {\enquote
  {\bibinfo {title} {Multiscale-structuring of polyvinylidene fluoride for
  energy harvesting: the impact of molecular-{,} micro- and macro-structure},}\
  }\href {\doibase 10.1039/C6TA09590A} {\bibfield  {journal} {\bibinfo
  {journal} {J. Mater. Chem. A}\ }\textbf {\bibinfo {volume} {5}},\ \bibinfo
  {pages} {3091--3128} (\bibinfo {year} {2017})}\BibitemShut {NoStop}%
\bibitem [{\citenamefont {Sano}\ \emph {et~al.}(2020)\citenamefont {Sano},
  \citenamefont {Ataka}, \citenamefont {Hashiguchi},\ and\ \citenamefont
  {Toshiyoshi}}]{mi11030267}%
  \BibitemOpen
  \bibfield  {author} {\bibinfo {author} {\bibfnamefont {Chikako}\ \bibnamefont
  {Sano}}, \bibinfo {author} {\bibfnamefont {Manabu}\ \bibnamefont {Ataka}},
  \bibinfo {author} {\bibfnamefont {Gen}\ \bibnamefont {Hashiguchi}}, \ and\
  \bibinfo {author} {\bibfnamefont {Hiroshi}\ \bibnamefont {Toshiyoshi}},\
  }\bibfield  {title} {\enquote {\bibinfo {title} {An electret-augmented
  low-voltage mems electrostatic out-of-plane actuator for acoustic transducer
  applications},}\ }\href {\doibase 10.3390/mi11030267} {\bibfield  {journal}
  {\bibinfo  {journal} {Micromachines}\ }\textbf {\bibinfo {volume} {11}},\
  \bibinfo {pages} {267} (\bibinfo {year} {2020})}\BibitemShut {NoStop}%
\bibitem [{\citenamefont {Jean-Mistral}\ \emph {et~al.}(2012)\citenamefont
  {Jean-Mistral}, \citenamefont {Vu~Cong},\ and\ \citenamefont
  {Sylvestre}}]{Jean-Mistral2012}%
  \BibitemOpen
  \bibfield  {author} {\bibinfo {author} {\bibfnamefont {C.}~\bibnamefont
  {Jean-Mistral}}, \bibinfo {author} {\bibfnamefont {T.}~\bibnamefont
  {Vu~Cong}}, \ and\ \bibinfo {author} {\bibfnamefont {A.}~\bibnamefont
  {Sylvestre}},\ }\bibfield  {title} {\enquote {\bibinfo {title} {Advances for
  dielectric elastomer generators: Replacement of high voltage supply by
  electret},}\ }\href {\doibase 10.1063/1.4761949} {\bibfield  {journal}
  {\bibinfo  {journal} {Applied Physics Letters}\ }\textbf {\bibinfo {volume}
  {101}},\ \bibinfo {pages} {162901} (\bibinfo {year} {2012})},\ \Eprint
  {http://arxiv.org/abs/https://doi.org/10.1063/1.4761949}
  {https://doi.org/10.1063/1.4761949} \BibitemShut {NoStop}%
\bibitem [{\citenamefont {Gross}\ and\ \citenamefont
  {de~Moraes}(1962)}]{Gross62}%
  \BibitemOpen
  \bibfield  {author} {\bibinfo {author} {\bibfnamefont {Bernhard}\
  \bibnamefont {Gross}}\ and\ \bibinfo {author} {\bibfnamefont {R.~J.}\
  \bibnamefont {de~Moraes}},\ }\bibfield  {title} {\enquote {\bibinfo {title}
  {Polarization of the electret},}\ }\href {\doibase 10.1063/1.1733151}
  {\bibfield  {journal} {\bibinfo  {journal} {The Journal of Chemical Physics}\
  }\textbf {\bibinfo {volume} {37}},\ \bibinfo {pages} {710--713} (\bibinfo
  {year} {1962})},\ \Eprint
  {http://arxiv.org/abs/https://doi.org/10.1063/1.1733151}
  {https://doi.org/10.1063/1.1733151} \BibitemShut {NoStop}%
\bibitem [{\citenamefont {Zhao}\ \emph {et~al.}(2020)\citenamefont {Zhao},
  \citenamefont {Goryachev}, \citenamefont {Krupka},\ and\ \citenamefont
  {Tobar}}]{Zhao2020}%
  \BibitemOpen
  \bibfield  {author} {\bibinfo {author} {\bibfnamefont {Zijun~C.}\
  \bibnamefont {Zhao}}, \bibinfo {author} {\bibfnamefont {Maxim}\ \bibnamefont
  {Goryachev}}, \bibinfo {author} {\bibfnamefont {Jerzy}\ \bibnamefont
  {Krupka}}, \ and\ \bibinfo {author} {\bibfnamefont {Michael~E.}\ \bibnamefont
  {Tobar}},\ }\bibfield  {title} {\enquote {\bibinfo {title} {Emergence of
  dielectric anisotropy of crystalline strontium titanate due to
  temperature-dependent phase transitions},}\ }\href@noop {} {\bibfield
  {journal} {\bibinfo  {journal} {arXiv:2008.07088 [cond-mat.mtrl-sci]}\ }
  (\bibinfo {year} {2020})}\BibitemShut {NoStop}%
\bibitem [{\citenamefont {Liu}\ \emph {et~al.}(2020)\citenamefont {Liu},
  \citenamefont {Tang}, \citenamefont {Li}, \citenamefont {Zhang},
  \citenamefont {Yu},\ and\ \citenamefont {Gu}}]{LIU2020}%
  \BibitemOpen
  \bibfield  {author} {\bibinfo {author} {\bibfnamefont {X.Z.}\ \bibnamefont
  {Liu}}, \bibinfo {author} {\bibfnamefont {Z.X.}\ \bibnamefont {Tang}},
  \bibinfo {author} {\bibfnamefont {Q.H.}\ \bibnamefont {Li}}, \bibinfo
  {author} {\bibfnamefont {Q.H.}\ \bibnamefont {Zhang}}, \bibinfo {author}
  {\bibfnamefont {X.Q.}\ \bibnamefont {Yu}}, \ and\ \bibinfo {author}
  {\bibfnamefont {L.}~\bibnamefont {Gu}},\ }\bibfield  {title} {\enquote
  {\bibinfo {title} {Symmetry-induced emergent electrochemical properties for
  rechargeable batteries},}\ }\href {\doibase
  https://doi.org/10.1016/j.xcrp.2020.100066} {\bibfield  {journal} {\bibinfo
  {journal} {Cell Reports Physical Science}\ }\textbf {\bibinfo {volume} {1}},\
  \bibinfo {pages} {100066} (\bibinfo {year} {2020})}\BibitemShut {NoStop}%
\bibitem [{\citenamefont {Xu}\ \emph {et~al.}(2019)\citenamefont {Xu},
  \citenamefont {Chien}, \citenamefont {Shi}, \citenamefont {Li}, \citenamefont
  {Wu}, \citenamefont {Liu}, \citenamefont {Hu},\ and\ \citenamefont
  {Goodenough}}]{Xu18815}%
  \BibitemOpen
  \bibfield  {author} {\bibinfo {author} {\bibfnamefont {Henghui}\ \bibnamefont
  {Xu}}, \bibinfo {author} {\bibfnamefont {Po-Hsiu}\ \bibnamefont {Chien}},
  \bibinfo {author} {\bibfnamefont {Jianjian}\ \bibnamefont {Shi}}, \bibinfo
  {author} {\bibfnamefont {Yutao}\ \bibnamefont {Li}}, \bibinfo {author}
  {\bibfnamefont {Nan}\ \bibnamefont {Wu}}, \bibinfo {author} {\bibfnamefont
  {Yuanyue}\ \bibnamefont {Liu}}, \bibinfo {author} {\bibfnamefont {Yan-Yan}\
  \bibnamefont {Hu}}, \ and\ \bibinfo {author} {\bibfnamefont {John~B.}\
  \bibnamefont {Goodenough}},\ }\bibfield  {title} {\enquote {\bibinfo {title}
  {High-performance all-solid-state batteries enabled by salt bonding to
  perovskite in poly(ethylene oxide)},}\ }\href {\doibase
  10.1073/pnas.1907507116} {\bibfield  {journal} {\bibinfo  {journal}
  {Proceedings of the National Academy of Sciences}\ }\textbf {\bibinfo
  {volume} {116}},\ \bibinfo {pages} {18815--18821} (\bibinfo {year} {2019})},\
  \Eprint
  {http://arxiv.org/abs/https://www.pnas.org/content/116/38/18815.full.pdf}
  {https://www.pnas.org/content/116/38/18815.full.pdf} \BibitemShut {NoStop}%
\bibitem [{\citenamefont {Ahmad}\ \emph {et~al.}(2018)\citenamefont {Ahmad},
  \citenamefont {George}, \citenamefont {Beesley}, \citenamefont {Baumberg},\
  and\ \citenamefont {De~Volder}}]{Ahmad:2018ro}%
  \BibitemOpen
  \bibfield  {author} {\bibinfo {author} {\bibfnamefont {Shahab}\ \bibnamefont
  {Ahmad}}, \bibinfo {author} {\bibfnamefont {Chandramohan}\ \bibnamefont
  {George}}, \bibinfo {author} {\bibfnamefont {David~J.}\ \bibnamefont
  {Beesley}}, \bibinfo {author} {\bibfnamefont {Jeremy~J.}\ \bibnamefont
  {Baumberg}}, \ and\ \bibinfo {author} {\bibfnamefont {Michael}\ \bibnamefont
  {De~Volder}},\ }\bibfield  {title} {\enquote {\bibinfo {title}
  {Photo-rechargeable organo-halide perovskite batteries},}\ }\bibfield
  {booktitle} {\emph {\bibinfo {booktitle} {Nano Letters}},\ }\href {\doibase
  10.1021/acs.nanolett.7b05153} {\bibfield  {journal} {\bibinfo  {journal}
  {Nano Letters}\ }\textbf {\bibinfo {volume} {18}},\ \bibinfo {pages}
  {1856--1862} (\bibinfo {year} {2018})}\BibitemShut {NoStop}%
\bibitem [{\citenamefont {Birch}(1985)}]{Birch_1985}%
  \BibitemOpen
  \bibfield  {author} {\bibinfo {author} {\bibfnamefont {C}~\bibnamefont
  {Birch}},\ }\bibfield  {title} {\enquote {\bibinfo {title} {The amperian
  current model of magnetisation and the prolate spheroid},}\ }\href {\doibase
  10.1088/0143-0807/6/3/010} {\bibfield  {journal} {\bibinfo  {journal}
  {European Journal of Physics}\ }\textbf {\bibinfo {volume} {6}},\ \bibinfo
  {pages} {180--182} (\bibinfo {year} {1985})}\BibitemShut {NoStop}%
\bibitem [{\citenamefont {Erturk}\ and\ \citenamefont
  {Inman}(2011)}]{Ertuk2011}%
  \BibitemOpen
  \bibfield  {author} {\bibinfo {author} {\bibfnamefont {Alper}\ \bibnamefont
  {Erturk}}\ and\ \bibinfo {author} {\bibfnamefont {Daniel~J.}\ \bibnamefont
  {Inman}},\ }\href@noop {} {\emph {\bibinfo {title} {PIEZOELECTRIC ENERGY
  HARVESTING}}}\ (\bibinfo  {publisher} {John Wiley and Sons, Ltd},\ \bibinfo
  {year} {2011})\BibitemShut {NoStop}%
\bibitem [{\citenamefont {Ikeda}(1996)}]{ikeda1996}%
  \BibitemOpen
  \bibfield  {author} {\bibinfo {author} {\bibfnamefont {T}~\bibnamefont
  {Ikeda}},\ }\href@noop {} {\emph {\bibinfo {title} {Fundamentals of
  Piezoelectricity}}}\ (\bibinfo  {publisher} {Oxford University Press, New
  York},\ \bibinfo {year} {1996})\BibitemShut {NoStop}%
\bibitem [{\citenamefont {Yang}(2018)}]{Yang2018}%
  \BibitemOpen
  \bibfield  {author} {\bibinfo {author} {\bibfnamefont {Jiashi}\ \bibnamefont
  {Yang}},\ }\href@noop {} {\emph {\bibinfo {title} {An Introduction to the
  Theory of Piezoelectricity}}}\ (\bibinfo  {publisher} {Springer Nature
  Switzerland},\ \bibinfo {year} {2018})\BibitemShut {NoStop}%
\bibitem [{\citenamefont {Dahiya}\ and\ \citenamefont
  {Valle}(2013)}]{Dahiya2013}%
  \BibitemOpen
  \bibfield  {author} {\bibinfo {author} {\bibfnamefont {R.S.}\ \bibnamefont
  {Dahiya}}\ and\ \bibinfo {author} {\bibfnamefont {M.}~\bibnamefont {Valle}},\
  }\href@noop {} {\emph {\bibinfo {title} {Robotic Tactile Sensing}}}\
  (\bibinfo  {publisher} {Springer Science+Business Media Dordrecht},\ \bibinfo
  {year} {2013})\BibitemShut {NoStop}%
\bibitem [{\citenamefont {Wang}(2020)}]{WANG2020104272}%
  \BibitemOpen
  \bibfield  {author} {\bibinfo {author} {\bibfnamefont {Zhong~Lin}\
  \bibnamefont {Wang}},\ }\bibfield  {title} {\enquote {\bibinfo {title} {On
  the first principle theory of nanogenerators from maxwell's equations},}\
  }\href {\doibase https://doi.org/10.1016/j.nanoen.2019.104272} {\bibfield
  {journal} {\bibinfo  {journal} {Nano Energy}\ }\textbf {\bibinfo {volume}
  {68}},\ \bibinfo {pages} {104272} (\bibinfo {year} {2020})}\BibitemShut
  {NoStop}%
\bibitem [{\citenamefont {Boisseau}\ \emph {et~al.}(2011)\citenamefont
  {Boisseau}, \citenamefont {Despesse}, \citenamefont {Ricart}, \citenamefont
  {Defay},\ and\ \citenamefont {Sylvestre}}]{Boisseau_2011}%
  \BibitemOpen
  \bibfield  {author} {\bibinfo {author} {\bibfnamefont {S}~\bibnamefont
  {Boisseau}}, \bibinfo {author} {\bibfnamefont {G}~\bibnamefont {Despesse}},
  \bibinfo {author} {\bibfnamefont {T}~\bibnamefont {Ricart}}, \bibinfo
  {author} {\bibfnamefont {E}~\bibnamefont {Defay}}, \ and\ \bibinfo {author}
  {\bibfnamefont {A}~\bibnamefont {Sylvestre}},\ }\bibfield  {title} {\enquote
  {\bibinfo {title} {Cantilever-based electret energy harvesters},}\ }\href
  {\doibase 10.1088/0964-1726/20/10/105013} {\bibfield  {journal} {\bibinfo
  {journal} {Smart Materials and Structures}\ }\textbf {\bibinfo {volume}
  {20}},\ \bibinfo {pages} {105013} (\bibinfo {year} {2011})}\BibitemShut
  {NoStop}%
\bibitem [{\citenamefont {{Suzuki}}(2015)}]{Suzuki2015}%
  \BibitemOpen
  \bibfield  {author} {\bibinfo {author} {\bibfnamefont {Y.}~\bibnamefont
  {{Suzuki}}},\ }\bibfield  {title} {\enquote {\bibinfo {title} {Electret based
  vibration energy harvester for sensor network},}\ }in\ \href {\doibase
  10.1109/TRANSDUCERS.2015.7180856} {\emph {\bibinfo {booktitle} {2015
  Transducers - 2015 18th International Conference on Solid-State Sensors,
  Actuators and Microsystems (TRANSDUCERS)}}}\ (\bibinfo {year} {2015})\ pp.\
  \bibinfo {pages} {43--46}\BibitemShut {NoStop}%
\bibitem [{\citenamefont {Kashiwagi}\ \emph {et~al.}(2011)\citenamefont
  {Kashiwagi}, \citenamefont {Okano}, \citenamefont {Miyajima}, \citenamefont
  {Sera}, \citenamefont {Tanabe}, \citenamefont {Morizawa},\ and\ \citenamefont
  {Suzuki}}]{Kashiwagi_2011}%
  \BibitemOpen
  \bibfield  {author} {\bibinfo {author} {\bibfnamefont {Kimiaki}\ \bibnamefont
  {Kashiwagi}}, \bibinfo {author} {\bibfnamefont {Kuniko}\ \bibnamefont
  {Okano}}, \bibinfo {author} {\bibfnamefont {Tatsuya}\ \bibnamefont
  {Miyajima}}, \bibinfo {author} {\bibfnamefont {Yoichi}\ \bibnamefont {Sera}},
  \bibinfo {author} {\bibfnamefont {Noriko}\ \bibnamefont {Tanabe}}, \bibinfo
  {author} {\bibfnamefont {Yoshitomi}\ \bibnamefont {Morizawa}}, \ and\
  \bibinfo {author} {\bibfnamefont {Yuji}\ \bibnamefont {Suzuki}},\ }\bibfield
  {title} {\enquote {\bibinfo {title} {Nano-cluster-enhanced high-performance
  perfluoro-polymer electrets for energy harvesting},}\ }\href {\doibase
  10.1088/0960-1317/21/12/125016} {\bibfield  {journal} {\bibinfo  {journal}
  {Journal of Micromechanics and Microengineering}\ }\textbf {\bibinfo {volume}
  {21}},\ \bibinfo {pages} {125016} (\bibinfo {year} {2011})}\BibitemShut
  {NoStop}%
\bibitem [{\citenamefont {Kudryavtsev}\ and\ \citenamefont
  {Trashkeev}(2013)}]{Kudryavtsev2013}%
  \BibitemOpen
  \bibfield  {author} {\bibinfo {author} {\bibfnamefont {A.~N.}\ \bibnamefont
  {Kudryavtsev}}\ and\ \bibinfo {author} {\bibfnamefont {S.~I.}\ \bibnamefont
  {Trashkeev}},\ }\bibfield  {title} {\enquote {\bibinfo {title} {Formalism of
  two potentials for the numerical solution of maxwell's equations},}\ }\href
  {\doibase 10.1134/S0965542513110079} {\bibfield  {journal} {\bibinfo
  {journal} {Computational Mathematics and Mathematical Physics}\ }\textbf
  {\bibinfo {volume} {53}},\ \bibinfo {pages} {1653--1663} (\bibinfo {year}
  {2013})}\BibitemShut {NoStop}%
\bibitem [{\citenamefont {Cabibbo}\ and\ \citenamefont
  {Ferrari}(1962)}]{Cabibbo1962}%
  \BibitemOpen
  \bibfield  {author} {\bibinfo {author} {\bibfnamefont {N.}~\bibnamefont
  {Cabibbo}}\ and\ \bibinfo {author} {\bibfnamefont {E.}~\bibnamefont
  {Ferrari}},\ }\bibfield  {title} {\enquote {\bibinfo {title} {Quantum
  electrodynamics with dirac monopoles},}\ }\href {\doibase 10.1007/BF02731275}
  {\bibfield  {journal} {\bibinfo  {journal} {Il Nuovo Cimento (1955-1965)}\
  }\textbf {\bibinfo {volume} {23}},\ \bibinfo {pages} {1147--1154} (\bibinfo
  {year} {1962})}\BibitemShut {NoStop}%
\bibitem [{\citenamefont {Keller}(2018)}]{Keller2018}%
  \BibitemOpen
  \bibfield  {author} {\bibinfo {author} {\bibfnamefont {Ole}\ \bibnamefont
  {Keller}},\ }\bibfield  {title} {\enquote {\bibinfo {title} {Electrodynamics
  with magnetic monopoles: Photon wave mechanical theory},}\ }\href {\doibase
  10.1103/PhysRevA.98.052112} {\bibfield  {journal} {\bibinfo  {journal} {Phys.
  Rev. A}\ }\textbf {\bibinfo {volume} {98}},\ \bibinfo {pages} {052112}
  (\bibinfo {year} {2018})}\BibitemShut {NoStop}%
\bibitem [{\citenamefont {Asker}(2018)}]{Asker2018}%
  \BibitemOpen
  \bibfield  {author} {\bibinfo {author} {\bibfnamefont {Andreas}\ \bibnamefont
  {Asker}},\ }\href@noop {} {\emph {\bibinfo {title} {Axion Electrodynamics and
  Measurable Effects in Topological Insulators}}}\ (\bibinfo  {publisher}
  {Kaerstads University},\ \bibinfo {year} {2018})\BibitemShut {NoStop}%
\bibitem [{\citenamefont {Tobar}\ \emph {et~al.}(2021)\citenamefont {Tobar},
  \citenamefont {Chiao},\ and\ \citenamefont {Goryachev}}]{Tobar2021}%
  \BibitemOpen
  \bibfield  {author} {\bibinfo {author} {\bibfnamefont {Michael~E.}\
  \bibnamefont {Tobar}}, \bibinfo {author} {\bibfnamefont {Raymond~Y.}\
  \bibnamefont {Chiao}}, \ and\ \bibinfo {author} {\bibfnamefont {Maxim}\
  \bibnamefont {Goryachev}},\ }\bibfield  {title} {\enquote {\bibinfo {title}
  {Dual aharanov-bohm berry phase due to the generation of electricity through
  permanent bound and free charge polarization},}\ }\href@noop {} {\bibfield
  {journal} {\bibinfo  {journal} {arXiv:2101.00945 [physics.class-ph]}\ }
  (\bibinfo {year} {2021})}\BibitemShut {NoStop}%
\end{thebibliography}
\end{document}